\documentclass{fundam}

\usepackage[utf8]{inputenc} 
\usepackage[T1]{fontenc} 
\usepackage{amsmath,amssymb}
\usepackage[pdftex]{graphicx}
\usepackage[all]{xy}
\usepackage{rotating}
\usepackage{hyperref} 
\usepackage{graphicx} 
\usepackage{enumerate} 
\usepackage{xcolor} 

\allowdisplaybreaks

\newcommand \RelView  {{\sc Rel\-View}}        

\newcommand \NAT      {{\mathbb N}}

\newcommand \REAL     {{\mathbb R}}
\newcommand \COMPL    {{\mathbb C}}
\newcommand \dom[1]   {\mbox{dom}(#1)}
\newcommand \inj      {\mbox{inj}}

\newcommand \I        {{\sf I}}

\newcommand \Comp     {;}
\newcommand \Transp[1]{{#1}^{\sf T}}

\newcommand \FP       {\textsl{fp}}
\newcommand \RP       {\textsl{rp}}
\newcommand \INV      {\textsl{J}}

\newcommand{\transp}[2]{{<}#1,#2{>}}

\renewcommand{\Comp}{\mathord{;}}

\renewcommand{\d}{\mathrm{d}}

\begin{document}

\title{Counting Specific Classes of Relations\\
Regarding Fixed Points and Reflexive Points}

\author{Rudolf Berghammer \\
        Institut für Informatik \\
        Christian-Albrechts-Universität zu Kiel \\
        24098 Kiel, Germany \\
        \texttt{rub@informatik.uni-kiel.de}
        \and
        Jules Desharnais\\
        Départment d'informatique et de génie logiciel \\
        Université Laval \\
        Québec, QC, Canada, G1V 0A6 \\
        \texttt{jules.desharnais@ift.ulaval.ca}
        \and
        Michael Winter\\
        Department of Computer Science\\
        Brock University \\
        St.\ Catharines, Ontario, Canada, L2S 3A1 \\
        \texttt{mwinter@brocku.ca}
       }

\date{Received: date / Revised: date}

\runninghead{R.\ Berghammer, J.\ Desharnais, M.\ Winter}
            {Counting Specific Classes of Relations}

\maketitle

\begin{abstract}
Given a finite and non-empty set $X$
and randomly selected specific functions and relations on $X$,
we investigate the existence and non-existence of fixed points and reflexive points, respectively.
First, we consider the class of functions, weaken it to the classes of 
partial functions, total relations and general relations and 
also strengthen it to the class of permutations.
Then we investigate the class of involutions and the subclass of 
proper involutions.
Finally, we treat idempotent functions, partial idempotent 
functions and related concepts.
We count relations, calculate corresponding probabilities and also 
calculate the limiting values of the latter in case that the 
cardinality of $X$ tends to infinity.
All these results have been motivated and also supported by numerous
experiments performed with the \RelView\ tool.
\end{abstract}


\section{Introduction}
\label{INTR} 

Fixed points of functions play a central role in mathematics and 
computer science.
Thus, a series of fixed point theorems have been developed that ensure 
their existence under certain conditions.
Well-known examples are the Knaster-Tarski fixed point theorem \cite{Knaster,Tarski} on 
complete lattices, the Kleene and Bourbaki-Witt fixed point theorems \cite{Bourbaki} on 
chain-complete posets, the Banach fixed point theorem on metric spaces \cite{Banach}, 
the Brouwer fixed point theorem on unit balls of Euclidean spaces \cite{Laborgne} and 
its generalization to topological Hausdorff vector spaces, called
Schauder fixed point theorem \cite{Bonsall}.

A question that naturally arises in this context concerns the portion 
of functions that have fixed points if no specific condition on the carrier
set is assumed.
Or, in other words:
Given a finite set $X$ with $n$ elements (or an $n$-set for short), 
where $n \in \NAT_{\geq 1} := \NAT \setminus\{0\}$, what is 
the probability that a 
randomly selected function on $X$ has a fixed point and what 
happens if $n$ tends to infinity?
To answer this question, we have considered functions $f : X \to X$ as 
univalent and total relations $f \subseteq X \times X$.
Then $f$ has a fixed point iff $f$ is not irreflexive, i.e., there exists
an element $x \in X$ such that $(x,x) \in f$.
Following \cite{DesMoe}, we call an element with this property also a
reflexive point of a relation.
Using the relational specifications $\mathfrak{unival}(X,Y)$,
$\mathfrak{total}(X,Y)$ and $\mathfrak{irrefl}(X)$ of 
\cite{Berghammer2021} for the sets of univalent and total 
relations between sets $X$ and $Y$ and the set of irreflexive relations 
on a set $X$, respectively, and their implementations as programs of 
the \RelView\ tool (cf. \cite{BergNeu,RELVIEW,RELVIEWGITHUB}), we have 
performed many experiments.
In \RelView\ relations are implemented via binary decision diagrams.
See \cite{BerLeoMil,BergNeu} for more details.
This very efficient implementation is particularly suitable
for the enumeration of huge sets of relations as, for example,
demonstrated in
\cite{BerMPC,Berghammer2021,BerKernel,BerLeoMil,BergNeu}.
In particular, we have been able to compute on $n$ elements all functions 
having a fixed point up to $n = 250$.
The results are shown in Figure \ref{fig:FUN-FIX}.
\begin{figure}[ht]
\begin{center}
$
\begin{array}{r|r|r|r}
n & \multicolumn{1}{c|}{\#\mbox{ functions}} & \multicolumn{1}{c|}{\#\mbox{ with fixed point}} & \multicolumn{1}{c}{\mbox{ratio}}\\
\hline
  1& 1.000000 \cdot   10^{0}& 1.000000 \cdot   10^{0}& 100.000 \% \\
  5& 3.125000 \cdot   10^{3}& 2.101000 \cdot   10^{3}&  67.232 \% \\
 10& 1.000000 \cdot  10^{10}& 6.513216 \cdot  10^{09}&  65.132 \% \\
 25& 8.881784 \cdot  10^{34}& 5.680818 \cdot  10^{34}&  63.960 \% \\
 50& 8.881784 \cdot  10^{84}& 5.647308 \cdot  10^{84}&  63.583 \% \\
100& 1.000000 \cdot 10^{200}& 6.339677 \cdot 10^{199}&  63.397 \% \\
150& 2.592321 \cdot 10^{326}& 1.641847 \cdot 10^{326}&  63.385 \% \\
200& 1.606938 \cdot 10^{460}& 1.017260 \cdot 10^{460}&  63.304 \% \\
250& 3.054936 \cdot 10^{599}& 1.933340 \cdot 10^{599}&  63.286 \% \\
\end{array}
$
\end{center}
\caption{Number of functions and functions with fixed points}
\label{fig:FUN-FIX}
\end{figure}
Looking at the table of Figure \ref{fig:FUN-FIX}, it seems that 
the probability that a randomly selected function on an
$n$-set converges to a number near to 0.6328.
In fact, it converges to $1 - \frac{1}{\rm e} \approx 0.6321205$,
as we will show in Section \ref{WITHFP}.

Inspired by this result, we have investigated
reflexive points (fixed points in the case of functions)
in the case of a randomly selected 
relation from a certain class of relations on an $n$-set in 
more detail.
In Section \ref{WITHFP} we concentrate on the existence
of fixed points. 
Besides functions, i.e., univalent and total relations as mentioned 
above, we also consider relations where only one or even none of 
these properties holds.  
That is, we also calculate the probability that a randomly 
selected partial function, total relation or arbitrary relation, 
respectively, on a set $X$ with $n$ elements has a fixed 
point respectively a reflexive point, and its limiting value 
if $n$ tends to infinity. 
In addition, we strengthen the assumption and answer the analog 
question for the set of permutations on $X$.

In each case the decisive part for obtaining the probability
consists in counting those relations of the class
under consideration which have at least one
reflexive point (fixed point for functions).
In Section \ref{WITHFP} we generalize these counting problems 
in two ways.
With the exception of permutations, first, we allow relations 
between possibly different sets, i.e., heterogeneous relations.
The only restriction is that the source is a subset of the
target.
This allows to define fixed points and reflexive points
also for such relations in an obvious way.
Second, we introduce an additional parameter $k$ and compute
for each of the five classes mentioned above the number of 
relations with $k$ fixed points respectively $k$ reflexive points.
So, for each class we obtain a list of numbers indicating the
number of relations of the class which have 0,1,2, \ldots\ fixed 
points respectively reflexive points.
For each of the five lists we then show how its members are 
related to each other with respect to their sizes.
To simplify presentation, for these comparisons we assume 
that source and target are equal, i.e., the relations are 
homogeneous.

There are some classes of relations which are defined by means of the
absence of reflexive points (fixed points for functions).
For instance, the irreflexive relations constitute such a class.
Also derangements are defined this way, viz. as permutations
without a fixed point.
In Section \ref{WITHFP} we will use derangements to determine
the number of permutations on an $n$-set with $k$ fixed 
points.
A further such class is that of proper involutions.
These are involutions (i.e., self-inverse permutations) without a
fixed point.
In Section \ref{WITHOUTFP} we compute the number of involutions
on an $n$-set $X$ that have $k$ fixed points.
This not only yields the number of proper involutions on $X$
but also allows to compute the probability that a randomly 
selected involution on $X$ is proper and its limiting value 
if $n$ tends to infinity.

A class of functions that always have fixed points is given by
the idempotent functions on a set $X$, since for such a function 
$f : X \to X$ each image element $f(x)$ is a fixed point of $f$.
If totality of $f$ is not demanded, then $f$ becomes an idempotent 
partial function on $X$.
In such a case it may happen that $f$ has no fixed point.
Due to this fact, in Section \ref{IDEMP} we investigate 
idempotent functions, idempotent partial functions and 
their weakening to transitive partial functions on $n$-sets 
and present solutions of problems similar to those we have 
treated in Sections \ref{WITHFP} and \ref{WITHOUTFP}.

Section \ref{CONCL} contains some concluding remarks.
In this section we also discuss related work and topics
for future investigations.
Especially, we describe an open problem concerning the probability of 
a randomly selected relation on an $n$-set to have a kernel.


\section{Preliminaries}
\label{PRELIM}

Relations are sets and, therefore, we can form the union $R \cup S$,
intersection $R \cap S$ and difference $R \setminus S$ of two relations 
$R$ and $S$.
In this context the empty set $\emptyset$ is called the empty relation.
That relations are sets of pairs is used in case of transposition
and composition, where the transpose $\Transp{R} \subseteq Y \times X$ 
of a relation $R \subseteq X \times Y$ is defined as
$\Transp{R} := \{(y,x) \mid (x,y) \in R\}$ and the composition 
$R \Comp S \subseteq X \times Z$ of a relation 
$R \subseteq X \times Y$ and a relation $S \subseteq Y \times Z$ 
is defined as
$R \Comp S := \{(x,z) \mid \exists\, y : (x,y) \in R \wedge (y,z) \in S\}$.
For a set $A$ the identity relation $\I_A$ on $A$ is defined as
$\I_A := \{(x,x) \mid x \in A\}$.
In the introduction we already have mentioned univalency, totality
and irreflexivity as specific properties of relations.
We assume that the reader is familiar with these notions and with 
the notions of reflexivity, symmetry and transitivity which we 
also will use throughout the paper.
As already mentioned, functions are univalent and total relations.
Univalent relations are also called partial functions.
As usual, in case of a (partial) function $f$
we write $f : X \to Y$ instead of $f \subseteq X \times Y$ 
and also $f(x) = y$ instead of $(x,y) \in f$. 
Furthermore, by $\dom{f}$ we denote the domain of $f$, i.e., the set 
of all $x \in X$ such that there exists a $y \in Y$ with $f(x) = y$. 
For more information concerning relations we refer to the textbook
\cite{Schmidt2011}.

There is a 1-1 correspondence between the set of relations from $X$ to $Y$
and the set of functions from $X$ to $2^Y$, where $2^Y$ denotes the 
powerset of $Y$, i.e., the set of subsets of $Y$. 
The bijection is given by $R \mapsto f_R$ where $f_R : X \to 2^Y$ 
is defined by 
$
f_R(x) = \{y \in Y \mid (x,y) \in R\},
$
for all $x \in X$. 
Furthermore, $R$ is (left) total, i.e., for all $x\in X$ there is a $y\in Y$ so that $(x,y)\in R$, iff the empty set is not in the image of $f_R$.
In this case we define $f^\ast_R : X \to 2^Y \setminus \{\emptyset\}$ to 
be the function $f_R$ restricted to non-empty subsets of $Y$.

If $f : X \to Y$ is a partial function, with $\bot \not\in Y$ then 
$f^t : X \to Y \cup\{\bot\}$ is a function if we define
$f^t(x) = f(x)$, for all $x \in \dom{f}$, and
$f^t(x) = \bot$, for all $x \in X \setminus \dom{f}$.
In fact, $f \mapsto f^t$ describes a bijection between the set of 
partial functions from $X$ to $Y$ and the set of functions from
$X$ to $Y \cup \{\bot\}$.

For our investigations it turned out to be convenient to work with a 
slightly generalized notion of a fixed point by moving from (partial) 
functions on a single set $X$ to (partial) functions from $X$ to a 
superset $Y$ of $X$. 
The definition of a fixed point for this generalization is obvious. 
Similarly, we also generalize the notion of a reflexive point from relations 
on a single set $X$ to relations with $X$ as source and a superset $Y$ 
of $X$ as target.

By means of the factorial $n!$ of natural numbers $n$, the subfactorial 
$!n$ of a natural number $n$ is defined as 
$
!n := n! \sum\limits_{k=0}^n \frac{(-1)^k}{k!}. 
$
Then the number of permutations on a set with $n$ elements 
without a fixed point
(such permutations are also called derangements) is exactly $!n$; see 
\cite{Stanley2012}, for instance.
In addition, we present some basic facts concerning subfactorials,
where in part (b) of Lemma~\ref{SUBFAC} the floor of a non-negative real number
$a$ is used, defined as
$\lfloor a \rfloor := \max \{ n \in \NAT \mid n \leq a\}$,
and {\rm e} is the base of the natural logarithm, that is,
${\rm e} := \sum\limits_{n = 0}^{\infty} \frac{1}{n!}$.

\begin{lemma}
\label{SUBFAC}
For all $n \in \NAT_{\geq 1}$ we have:
\begin{enumerate}[(a)]
\item[(a)] $!n =n\cdot~!(n-1) + (-1)^n$.
\item[(b)] $!n = \lfloor \frac{n! + 1}{{\rm e}} \rfloor$.
\end{enumerate}
\end{lemma}

\noindent
For part (a) see \cite{Stanley2012} and for part (b) see \cite{WeissteinSf}.
Part (a) in combination with $!0 = 1$ allows a recursive computation of 
subfactorials.
{From} the lemma we also get that the sequence $(!n)_{n \geq 1}$
is strictly increasing, starting with 
$!1$ $=$ $1$ $\cdot$ $!0$ $-1$ $=$ $0$, and that
$!n > 0$ for all $n \in \mathbb{N}_{\geq 2}$.

We assume the reader also to be familiar with the binomial coefficient
$\binom{n}{k}$ of $n, k \in \NAT$ with $k \leq n$, defined by
$ 
\binom{n}{k} := \frac{n!}{k!(n-k)!}. 
$
A well-known result is that
$(x+y)^n = \sum\limits_{k=0}^n \binom{n}{k} x^k y^{n-k}$,
for all $n \in \NAT$ and $x,y \in \REAL$.
This is called the binomial theorem.
For $x = y = 1$ it yields $\sum\limits_{k=0}^n \binom{n}{k} = 2^n$.
We also will use that $|2^X_k| = \binom{|X|}{k}$, for all finite sets $X$ 
and all $k \in \NAT$, where 
$2^X_k := \{ A \in 2^X \mid k = |A|\}$
denotes the set of $k$-subsets of $X$. 
Finally, we will use the 
following lemma the proof of which can be found in the appendix.

\begin{lemma}
\label{BINOM}
For all $n, k \in \NAT$ with $k < n$ we have:
\begin{enumerate}[(a)]
\item $\binom{n}{k+1} = \frac{n-k}{k+1} \binom{n}{k}$.
\item $\binom{n+1}{k} = \frac{n+1}{n+1-k} \binom{n}{k}$.
\item $\binom{n+1}{k+1} = \frac{n+1}{k+1} \binom{n}{k}$.
\end{enumerate}
\end{lemma}

\noindent
Also the following auxiliary results will be used later.
Lemma \ref{LEM1} states well-known facts about the number {\rm e};
the proof of the Lemmas \ref{LEMMW-1} and \ref{LEM2} again 
can be found in the
appendix.

\begin{lemma}
\label{LEM1}
We have:
\begin{enumerate}
\item[(a)] 
  $\lim\limits_{n \to \infty} \left (1 + \frac{1}{n} \right )^n = {\rm e}$.
\item[(b)] 
  $\lim\limits_{n \to \infty} 
   \left (1 - \frac{1}{n} \right )^n = \frac{1}{\rm e}$.
\item[(c)] 
  $\sum\limits_{k=0}^\infty \frac{(-1)^k}{k!} = \frac{1}{\rm e}$.
\end{enumerate}
\end{lemma}

\begin{lemma}
\label{LEMMW-1}
For all $n \in \NAT$ with $n \geq 8$ and $k \leq n$ 
we have $2^n k^{n-k} 2^k \leq n^{n-1}$.
\end{lemma}

\noindent
The following lemma uses the Lambert function
$W$, which is 
a multivalued function consisting of the branches of the 
converse relation of the complex function 
$f  : \COMPL \to \COMPL$, where $f(w) = w{\rm e}^w$, for all
$w \in \COMPL$. 
If we restrict the argument $w$ in $f$ to non-negative real 
numbers, then $W$ (branch $W_0$) is a real function that 
satisfies $w{\rm e}^w = z$ iff $w = W(z)$, for all 
$w, z \in \REAL_{\geq 0}$;
see \cite{WikiLambert}.

\begin{lemma}
\label{LEM2}
Define for every $n\geq 1$ a function 
$h_n : \REAL_{\geq 1} \to \REAL_{\geq 1}$ on the real numbers 
greater than or equal to $1$ by $h_n(k)=k^{n-k}$, for all
$k \in \REAL_{\geq 1}$.
Then we have:
\begin{enumerate}[(a)]
\item Each $h_n$ has its maximum at $k_n := {\rm e}^{W(n{\rm e})-1}
                                          = \frac{n}{W(n{\rm e})}$ 
       and is increasing in the interval $[1,k_n]$.
\item For all $n \geq 256$ we have 
      $2\lfloor 3\log_2(n)\rfloor+1\leq k_n$.
\end{enumerate}
\end{lemma}


\section{Relations with Fixed Points and Reflexive Points}
\label{WITHFP}

Assume $X$ to be a finite set with $m > 0$ elements.
In the first part of this section, we are going to count the number 
of functions, partial functions, total relations and arbitrary relations 
from $X$ to a finite superset $Y$ of $X$ that have exactly $k$ fixed points 
respectively $k$ reflexive points, where $k \in \{0,\ldots,m\}$.
Based upon this, we then count the number of elements of these four
classes of relations that have at least one fixed point respectively at least one 
reflexive point. 
This allows to calculate the probability that a randomly selected element 
from these classes has a fixed point respectively a reflexive point,
together with its limiting value.
In the second part of the section, we do the same for the class of
permutations on $X$.
Finally, in the third part we investigate for each of the five 
lists of numbers we have 
calculated in the first and second parts how their elements are related to
each other with respect to their sizes.
To simplify the presentation, for the computation of the limiting values
as well as for the comparisons of the third part we assume $X = Y$, i.e.,
that also the elements of the first four classes of relations are defined 
on a single set (i.e., are homogeneous relations).

The following theorem provides the number of functions, partial functions, 
total relations and arbitrary relations that have exactly $k$ fixed points 
respectively exactly $k$ reflexive points.

\begin{theorem}
\label{THEEXIS-GEN}
Assume $m, n \in \NAT_{\geq 1}$ and finite sets $X$ and $Y$
with $m = |X|$, $n = |Y|$ and $X \subseteq Y$.
Then we have for all $k \in \{0,\ldots,m\}$:
\begin{enumerate}
\item[(a)] The number of functions $f : X \to Y$ with $k$ fixed points is 
	$\binom{m}{k} (n-1)^{m-k}$.
\item[(b)] The number of partial functions $f : X \to Y$ with $k$ fixed points
	is $\binom{m}{k} n^{m-k}$.
\item[(c)] The number of total relations $R \subseteq X \times Y$ with 
        $k$ reflexive points is $\binom{m}{k} 2^{k(n-1)} (2^{n-1} - 1)^{m-k}$.
\item[(d)] The number of relations $R \subseteq X \times Y$ with 
       $k$ reflexive points is $\binom{m}{k} 2^{m(n-1)}$.
\end{enumerate}
\end{theorem}
\begin{proof}
(a) Any function $f : X \to Y$ can be split as $f = \inj_A \cup g$, 
where $A$ is the set of fixed points of $f$, the function 
$\inj_A : A \to Y$ is the canonical injection, i.e., $\inj_A(x) = x$, 
for all $x\in A$, and the function $g : X \setminus A \to Y$ has no 
fixed point. 
There are $\binom{m}{k}$ ways to select a set $A \in 2^X_k$. 
If $g : X \setminus A \to Y$ is a function without a fixed point, then 
for each $x \in X \setminus A$ there are precisely $n-1$ possible elements 
to which $x$ can be mapped to.
As a consequence, for each set $A \in 2^X_k$ there are $(n-1)^{m-k}$ ways to 
select a function $g : X \setminus A \to Y$ without a fixed point 
such that $f = \inj_A \cup g$.
Altogether, the number of functions $f : X \to Y$ with $k$ fixed points is
$\binom{m}{k} (n-1)^{m-k}$.

(b) A partial function $f : X\to Y$ has $k$ fixed points iff the 
function $f^t : X \to Y\cup\{\bot\}$ (defined in Section \ref{PRELIM})
has $k$ fixed points. 
Therefore, by (a) the number of partial functions $f : X \to Y$ 
with $k$ fixed points is
$\binom{m}{k} (n+1-1)^{m-k}=\binom{m}{k}n^{m-k}$.

(c) Let $R \subseteq X \times Y$ be a total relation, $A$ be the set 
of reflexive points of $R$ and assume that $|A| = k$.
We consider the function
$f^\ast_R : X \to 2^Y \setminus \{\emptyset\}$,
defined in Section \ref{PRELIM}.
Then, we have $x \in f^\ast_R(x)$, for all $x \in A$, and 
$x \notin f^\ast_R(x)$, for all $x \in X \setminus A$. 
For a given $x \in X$ there are $2^{n-1}-1$ non-empty subsets $M$ of 
$Y$ with $x \notin M$ and there are $2^{n-1}$ subsets $M$ of $Y$ 
with $x \in M$. 
This gives us $(2^{n-1}-1)^{m-k}(2^{n-1})^k$ functions 
$f : X \to 2^Y \setminus \{\emptyset\}$, for any given set $A$ of $k$ 
reflexive points.
Altogether, we obtain $\binom{m}{k} 2^{k(n-1)} (2^{n-1} - 1)^{m-k}$
total relations $R \subseteq X \times Y$ with $k$ reflexive points.

(d) Again, let $A$ be the set of reflexive points of a relation 
$R \subseteq X \times Y$, such that $|A| = k$.
Here we consider the function
$f_R : X \to 2^Y$ of Section \ref{PRELIM}.
Then we have $x \in f_R(x)$, for all $x\in A$, and $x \notin f_R(x)$,
for all $x \in X \setminus A$. 
For a given $x \in X$ there are $2^{n-1}$ subsets $M$ of $Y$ with 
$x \notin M$ and also $2^{n-1}$ subsets $M$ of $Y$ with $x \in M$. 
This gives us $(2^{n-1})^{m-k}(2^{n-1})^k = 2^{m(n-1)}$ functions 
$f : X \to 2^Y$, for any given set $A$ of $k$ reflexive points. 
Altogether, we obtain $\binom{m}{k}2^{m(n-1)}$ relations 
$R \subseteq X \times Y$ with $k$ reflexive points.
\end{proof}

\noindent
If we choose $k = 0$ in this theorem, then in combination
with $\binom{m}{0} = 1$, for all $m \in \NAT$, we immediately
obtain the following result.

\begin{corollary}
\label{LEM2GEN}
Assume $m, n \in \NAT_{\geq 1}$ and finite sets $X$ and $Y$
with $m = |X|$, $n = |Y|$ and $X \subseteq Y$.
Then we have:
\begin{enumerate}
\item[(a)] The number of functions $f : X \to Y$ without a fixed point 
           is $(n-1)^m$.
\item[(b)] The number of partial functions $f : X \to Y$ without 
           a fixed point is $n^m$.
\item[(c)] The number of total relations $R \subseteq X \times Y$ 
           without a reflexive point is $(2^{n-1}-1)^m$.
\item[(d)] The number of relations $R \subseteq X \times Y$ without a reflexive point 
           is $2^{m(n-1)}$.                      
\end{enumerate}
\end{corollary}

\noindent
Next, for each of the four classes of relations treated so far 
we calculate the number of elements that have at least one 
fixed point respectively at least one reflexive point and 
also the corresponding probability.
For the latter we assume that an element is chosen uniformly at random.

\begin{theorem}
\label{THEEXIS}
Assume $m, n \in \NAT_{\geq 1}$ and finite sets $X$ and $Y$
with $m = |X|$, $n = |Y|$ and $X \subseteq Y$.
Then we have
\begin{enumerate}
\item[(a)] The number of functions $f:X\to Y$ with a fixed point is $n^m - (n-1)^m$, 
      and the probability that a function $f:X\to Y$ selected uniformly at random 
      has a fixed point is $1-(\frac{n-1}{n})^m$.
\item[(b)] The number of partial functions $f:X\to Y$ with a fixed point is 
      $(n+1)^m - n^m$, and  the probability that a partial function $f:X\to Y$
      selected uniformly at random has a fixed point is $1 - ( \frac{n}{n+1} )^m$.
\item[(c)] The number of total relations $R\subseteq X\times Y$ with a reflexive point is 
      $(2^n-1)^m-(2^{n-1}-1)^m$, and the probability that a total relation
      $R\subseteq X\times Y$ selected uniformly at random has a reflexive point is 
      $1-(\frac{2^{n-1}-1}{2^n-1})^m$.
\item[(d)] The number of relations $R\subseteq X\times Y$ with a reflexive point is 
      $2^{mn}-2^{m(n-1)}$, and the probability that a relation $R\subseteq X\times Y$
      selected uniformly at random has a reflexive point is $1-\frac{1}{2^m}$.
\end{enumerate}
\end{theorem}
\begin{proof}
(a) There are $n^m$ functions from $X$ to $Y$.
By Corollary \ref{LEM2GEN}(a) exactly $(n-1)^m$ of them have no fixed point.
Hence, $n^m - (n-1)^m$ functions $f : X \to Y$ have a fixed point.
The probability that a function $f:X\to Y$ 
selected uniformly at random has a fixed point is calculated by
$$ 
\frac{n^m - (n-1)^m}{n^m} 
= 
1 - \frac{(n-1)^m}{n^m} 
= 
1 - \left ( \frac{n-1}{n} \right )^m. 
$$

(b) By the bijection $f\mapsto f^t$ stated in Section \ref{PRELIM} we know that there are
$(n+1)^m$ partial functions from $X$ to $Y$.
If we combine this with Corollary \ref{LEM2GEN}(b), then as in (a) 
we obtain the number of partial functions $f : X \to Y$ that have a
fixed point as $(n+1)^m - n^m$.
Again as in (a), the probability that a partial function 
$f : X \to Y$ selected uniformly at random has a fixed point is calculated
by
$$ 
\frac{(n+1)^m -n^m}{(n+1)^m} 
= 
1 - \frac{n^m}{(n+1)^m} 
= 
1 - \left ( \frac{n}{n+1} \right )^m. 
$$

(c) Also from Section \ref{PRELIM} we know that there exist
$(2^n-1)^m$ total relations $R \subseteq X \times Y$.
Due to Corollary \ref{LEM2GEN}(c) exactly
$(2^{n-1}-1)^m$ of them have no reflexive point and so
$(2^n-1)^m-(2^{n-1}-1)^m$ of them have a reflexive point.
Based upon this, the probability that a total relation $R \subseteq X \times Y$ 
selected uniformly at random has a reflexive point is calculated by
$$ 
\frac{(2^n-1)^m-(2^{n-1}-1)^m}{(2^n-1)^m} 
= 
1-\frac{(2^{n-1}-1)^m}{(2^n-1)^m} 
= 
1 - \left ( \frac{2^{n-1}-1}{2^n-1} \right )^m .
$$

(d) The number of relations $R \subseteq X \times Y$ is $2^{mn}$.
Corollary \ref{LEM2GEN}(d) shows that $2^{m(n-1)}$ of these relations
have no reflexive point.
As a consequence, $2^{mn}-2^{m(n-1)}$ relations $R \subseteq X \times Y$ 
have a reflexive point, and, hence, by the calculation
$$ 
\frac{2^{mn}-2^{m(n-1)}}{2^{mn}} 
= 
1-\frac{2^{m(n-1)}}{2^{mn}}
=
1-\frac{2^{mn}2^{-m}}{2^{mn}}
=
1-\frac{1}{2^m} 
$$
we get the probability that a relation $R \subseteq X \times Y$
selected uniformly at random has a reflexive point.
\end{proof}

\noindent
As already mentioned at the beginning of this section, we also will 
calculate the limiting values of the probabilities of the previous 
theorem in case that $X$ equals $Y$ and the cardinality of this
set tends to infinity.
The following theorem presents the results.

\begin{theorem}
\label{THEEXIS-limits}
Assume $n \in \NAT_{\geq 1}$ and a finite set $X$ 
with $n = |X|$.
If $n$ tends to infinity, then for the probabilities of 
Theorem \ref{THEEXIS} (using $X=Y$ and, hence, $m=n$)
we obtain the following limiting values:
\begin{enumerate}[(a)]
\item
 $\lim\limits_{n \to \infty} 1-\left(\frac{n-1}{n}\right)^n = 1-\frac{1}{\rm e}$.
\item
 $\lim\limits_{n \to \infty} 1-\left(\frac{n}{n+1} \right)^n = 1 - \frac{1}{\rm e}$.
\item
 $\lim\limits_{n \to \infty} 1 - \left ( \frac{2^{n-1} - 1}{2^n - 1} \right )^n = 1$.
\item
 $\lim\limits_{n \to \infty} 1-\frac{1}{2^n} = 1$.
\end{enumerate}
\end{theorem}
\begin{proof}
(a) We use the probability that a function selected uniformly at random 
has a fixed point from Theorem \ref{THEEXIS}(a)
and calculate
$$ 
\lim_{n \to \infty} 1-\left(\frac{n-1}{n}\right)^n
=
1-\lim_{n \to \infty} \left(1-\frac{1}{n} \right)^n
= 
1-\frac{1}{\rm e} ,
$$
where the last equality follows from Lemma \ref{LEM1}(b).

(b) This time we are using Theorem \ref{THEEXIS}(b).
We calculate
$$ 
\lim_{n \to \infty} 1-\left(\frac{n}{n+1} \right)^n 
= 
1-\lim_{n \to \infty} \frac{1}{\left ( \frac{n+1}{n} \right )^n} 
= 
1-\frac{1}{\lim\limits_{n \to \infty} \left ( 1 + \frac{1}{n} \right )^n} 
= 
1 - \frac{1}{\rm e},
$$
where now the last step uses Lemma \ref{LEM1}(a).

(c) Using Theorem \ref{THEEXIS}(c),
the convergence result is shown by
\[
   \lim_{n \to \infty} 1 - \left ( \frac{2^{n-1} - 1}{2^n - 1} \right )^n
= 1 - \lim_{n \to \infty} \left ( \frac{2^{n-1} - 1}{2^n - 1} \right )^n
= 1 - 0
= 1 .
\]
The second step of this calculation follows from
$\frac{2^{n-1} - 1}{2^n - 1} \leq \frac{1}{2}$,
for all $n \in \NAT_{\geq 1}$, and
$\lim\limits_{n \to \infty} ( \frac{1}{2} )^n = 0$.

(d) We use Theorem \ref{THEEXIS}(d)
and realize that
$\lim\limits_{n \to \infty} 1-\frac{1}{2^n} = 1$ is trivial. 
\end{proof}

\noindent
As fifth class of relations we now consider bijective functions.
Between finite sets a bijective function exists iff the
sets' cardinalities are equal.
As a consequence, we again assume $X = Y$.
A bijective function $f : X \to X$ is also called a permutation 
on $X$.
In the next theorems we consider these specific functions and
start again with the calculation of the number of those which have 
$k$ fixed points.
The following theorem is the counterpart to Theorem
\ref{THEEXIS-GEN}
for permutations, i.e., bijective functions on a finite set $X$.

\begin{theorem}
\label{THEPERM-GEN}
Assume $n \in \NAT_{\geq 1}$, a finite set $X$ with $n = |X|$
and $k \in \{0,\ldots,n\}$.
Then the number of permutations $f : X \to X$ with $k$ fixed points 
is $\binom{n}{k} \cdot~!(n-k)$.
\end{theorem}
\begin{proof}
As already used in the proof of Theorem \ref{THEEXIS-GEN}(a), any function 
$f : X \to X$ can be split as $f = \inj_A \cup g$, where $A$ is the set of 
fixed points of $f$, the function $\inj_A : A \to X$ is the 
canonical injection and the function $g : X \setminus A \to X$ 
has no fixed point. 
Next, we prove by contradiction that $f = \inj_A \cup g$ implies that 
the image of $g$ is a subset of $X \setminus A$.
For this, assume to the contrary that there exists $x \in X \setminus A$ 
such that $g(x) \in A$.
By the definition of $A$ this yields $f(g(x)) = g(x)$.
{From} $f = \inj_A \cup g$ and $x \in X \setminus A$ we get
$f(x) = g(x)$.
Altogether, we have $f(g(x)) = f(x)$, a contradiction to $f$ 
is injective since $g(x) \ne x$. 
As $f$ is assumed as bijective, $g$ is injective.
Because of $g(X \setminus A) \subseteq X \setminus A$,
we may restrict the target of $g$ to the set $X \setminus A$ 
and obtain this way $g$ as derangement on $X \setminus A$.

There are $\binom{m}{k}$ ways to select a set $A \in 2^X_k$. 
For each such $A$ there are $!(n-k)$ derangements
on $X \setminus A$ as already mentioned in Section \ref{PRELIM}
(see also \cite{Stanley2012}, Example 2.2.1.). 
Altogether, we obtain the number of permutations on $X$ 
with $k$ fixed points as 
$\binom{n}{k} \cdot~!(n-k)$.
\end{proof}

\noindent
We conclude the investigations concerning the existence of
fixed points of randomly selected specific relations 
with the following result on permutations.
Its first part is the counterpart to Theorem \ref{THEEXIS} and 
its second part is the counterpart to Theorem \ref{THEEXIS-limits}.

\begin{theorem}
\label{THEPERM}
Assume $n \in \NAT_{\geq 1}$ and a finite set $X$ with $n = |X|$.
Then we have:
\begin{enumerate}
\item[(a)]
 The number of permutations $f : X \to X$ with a fixed point is 
 $n!$ $-$ $!n$ and the probability that a permutation on $X$ selected uniformly
 at random has a fixed point is $1 - \sum\limits_{k=0}^n \frac{(-1)^k}{k!}$.
\item[(b)]
 If $n$ tends to infinity, then for the 
 probability of (a) we have
 $\lim\limits_{n \to \infty} 
 1 - \sum\limits_{k=0}^n \frac{(-1)^k}{k!} = 1 - \frac{1}{\rm e}$
 as limiting value.
\end{enumerate}
\end{theorem}
\begin{proof}
(a) There are $n!$ permutations on $X$.
In Section \ref{PRELIM} we have already mentioned that
the number of permutations on a set with $n$ elements without a fixed point
is $!n$.
So, the number of permutations on $X$ with a fixed point is 
$n!$ $-$ $!n$.
This implies that the probability that a permutation on $X$ selected 
uniformly at random has a fixed point is $\frac{n!\,-\,!n}{n!}$ 
and we obtain 
\[
  \frac{n!\,-\,!n}{n!} 
= 1 - \frac{!n}{n!} 
= 1 - \frac{n! \sum\limits_{k=0}^n \frac{(-1)^k}{k!}}{n!} 
= 1 - \sum_{k=0}^n \frac{(-1)^k}{k!} ,
\]
which yields the second claim.

(b) The calculation 
\[
  \lim_{n \to \infty} 1 - \sum_{k=0}^n \frac{(-1)^k}{k!}
= 1 - \lim_{n \to \infty} \sum_{k=0}^n \frac{(-1)^k}{k!}
= 1 - \sum_{k=0}^\infty \frac{(-1)^k}{k!}
= 1 - \frac{1}{\rm e}
\]
proves the claim. 
Here the last step uses equation (c) of Lemma \ref{LEM1}.
\end{proof}

\noindent
It is remarkable that three of the five probabilities we have
calculated in this section converge to the same limiting value
$1 - \frac{1}{\rm e}$.
Those of Theorem \ref{THEEXIS-limits}(a) converge ``from above'',
which is confirmed by the table in Figure \ref{fig:FUN-FIX}.
In case of Theorem \ref{THEPERM}(b) the probability is alternating.
For odd $n$ the probability also converges ``from above'',
but for even $n$ it converge ``from below'' to the limiting value 
$1-\frac{1}{e}$. This can also be seen from the values in the table of Figure \ref{fig:PERM-FIX}.
A comparison of the table of Figure \ref{fig:FUN-FIX} with the
table of Figure \ref{fig:PERM-FIX} shows that the second probability 
converges much faster than the first one. 

\begin{figure}[ht]
\begin{center}
$
\begin{array}{r|r|r|r}
n & \#\mbox{ permutations} & \#\mbox{ with fixed point} & \multicolumn{1}{c}{\mbox{ratio}}\\
\hline
  1 &             1 &            1 & 100.00000 \% \\
  2 &             2 &            1 &  50.00000 \% \\ 
  3 &             6 &            4 &  66.66666 \% \\
  4 &            24 &           15 &  62.50000 \% \\ 
  5 &           120 &           76 &  63.33333 \% \\
  6 &           720 &          455 &  63.19444 \% \\
  7 &          5040 &         3186 &  63.21428 \% \\
  8 &         40320 &        25487 &  63.21180 \% \\
  9 &        362880 &       229384 &  63.21208 \% \\
 10 &       3628800 &      2293839 &  63.21205 \% \\
 11 &      39916800 &     25232230 &  63.21205 \% \\
 12 &     479001600 &    302786759 &  63.21205 \% 
\end{array}
$
\end{center}
\caption{Number of permutations and permutations with fixed points}
\label{fig:PERM-FIX}
\end{figure}

The probability of Theorem \ref{THEEXIS-limits}(b) converges
``from below'' to the limiting value
$1 - \frac{1}{\rm e}$.
Experiments with the \RelView\ tool lead to the table of Figure
\ref{fig:PFUN-FIX}.
\begin{figure}[ht]
\begin{center}
$
\begin{array}{r|r|r|r}
n & \multicolumn{1}{c|}{\#\mbox{ partial functions}} & \multicolumn{1}{c|}{\#\mbox{ with fixed point}} & \multicolumn{1}{c}{\mbox{ratio}}\\
\hline
  1& 2.000000 \cdot   10^{0}& 1.000000 \cdot   10^{0}&  50.000 \% \\
  5& 7.776000 \cdot   10^{3}& 4.651000 \cdot   10^{3}&  59.812 \% \\
 10& 2.593742 \cdot  10^{10}& 1.593742 \cdot  10^{10}&  61.445 \% \\
 25& 2.367238 \cdot  10^{35}& 1.479560 \cdot  10^{35}&  62.501 \% \\
 50& 2.390610 \cdot  10^{85}& 1.502432 \cdot  10^{85}&  62.847 \% \\
100& 2.704814 \cdot 10^{200}& 1.704814 \cdot 10^{200}&  63.028 \% \\
150& 7.023314 \cdot 10^{326}& 4.430993 \cdot 10^{326}&  63.089 \% \\
200& 4.357240 \cdot 10^{460}& 2.750302 \cdot 10^{460}&  63.120 \% 
\end{array}
$
\end{center}
\caption{Number of partial functions and partial functions with fixed points}
\label{fig:PFUN-FIX}
\end{figure}
These numerical data show that the speed of convergence of the probability 
of Theorem \ref{THEEXIS-limits}(b) is similar to that of the probability 
of Theorem \ref{THEEXIS-limits}(a). 
This is in fact not surprising. 
Assuming $m = n$ and making the variable change of $n$ to 
$n + 1$ in the probability for functions (on a single $n$-set) 
given in Theorem \ref{THEEXIS}(a) yields $1-(\frac{n}{n+1})^{n+1}$.
This is almost the same as the probability 
$1-(\frac{n}{n+1})^n$ for partial functions
(on a single $n$-set) 
given in Theorem \ref{THEEXIS}(b).
The calculation
$
\textstyle
(1-(\frac{n}{n+1})^{n+1}) - (1-(\frac{n}{n+1})^n)
= (\frac{n}{n+1})^n - (\frac{n}{n+1})^{n+1}
= (\frac{n}{n+1})^n \frac{1}{n+1}
$
shows that the difference between the two quickly diminishes.

In the remainder of this section, we investigate how the members of
the five lists of numbers we have calculated in
Theorems \ref{THEEXIS-GEN} and \ref{THEPERM-GEN}
are related to each other with respect to their sizes.
To simplify the presentation, we assume $X = Y$ also for the four
classes treated in Theorem \ref{THEEXIS-GEN}, hence $m = n$ there.
So, all elements of the five classes of relations we have 
considered so far are defined on $n$-sets $X$ with $n \geq 1$.
For all $k \in \{0,\ldots,n\}$ we define 
\begin{enumerate}
\item[(1)]
  $\FP_n(k) := \binom{n}{k} (n-1)^{n-k}$ as the
  number of functions on $X$ that have $k$ fixed points,
\item[(2)]
  $\FP^u_n(k) := \binom{n}{k} n^{n-k}$ as the
  number of partial functions ($u$nivalent relations) on $X$ 
  that have $k$ fixed points,
\item[(3)]
  $\RP^t_n(k) := 
  \binom{n}{k} 2^{kn-k} (2^{n-1} - 1)^{n-k}$ as the
  number of $t$otal relations on $X$ that have $k$ reflexive points,
\item[(4)]
  $\RP_n(k) := 
  \binom{n}{k} 2^{n^2-n}$ as the
  number of relations on $X$ that have $k$ reflexive points, and
\item[(5)]
  $\FP^p_n(k) := \binom{n}{k} \cdot~!(n-k)$ as the
  number of $p$ermutations on $X$ that have $k$ fixed points.
\end{enumerate}
We start our investigations with the list
$\FP_n(0),\ldots,\FP_n(n)$ obtained by Theorem \ref{THEEXIS-GEN}(a).
Using Corollary \ref{LEM2GEN}(a) and $\binom{n}{1} = n$, we get
$
\FP_n(0) = (n-1)^n
      = (n-1) (n-1)^{n-1}
      < n (n-1)^{n-1}
      = \FP_n(1) .
$
This means that on $X$ there are strictly more functions with 
one fixed point than functions without a fixed point.
For a treatment of the sublist
$\FP_n(1),\ldots, \FP_n(n)$, we suppose $n \geq 2$
and use Lemma \ref{BINOM}(a) to calculate the following recursive 
description, where $k \in \{0,\ldots,n-1\}$:
\begin{align*}
  \FP_n(k+1)
&= \binom{n}{k+1} (n-1)^{n-(k+1)} \\
&= \frac{n-k}{k+1} \binom{n}{k} (n-1)^{n-k-1} \\
&= \frac{n-k}{(k+1)(n-1)} \binom{n}{k} (n-1)^{n-k} \\
&= \frac{n-k}{(k+1)(n-1)} \FP_n(k) 
\end{align*}
If we assume $k \geq 1$, then we get $\frac{n-k}{(k+1)(n-1)} < 1$ 
and the recursive description implies
$\FP_n(k+1) < \FP_n(k)$.
Using $k=1$ and then $k=0$, it yields
$
\FP_n(2) = \frac{n-1}{2(n-1)} \FP_n(1)
        = \frac{1}{2} \FP_n(1)
        = \frac{n}{2(n-1)} \FP_n(0)$.
So, in this case on $X$ there are twice as many functions with 
one fixed point than functions with two fixed points.
Furthermore, $n \geq 2$ implies $\frac{n}{2(n-1)} \leq 1$, hence 
$\FP_n(0) \geq \FP_n(2)$, where
$\FP_n(0) = \FP_n(2)$ iff $n=2$.
As a consequence $n \geq 3$ implies $\frac{n}{2(n-1)} < 1$ and we get
$\FP_n(0) > \FP_n(2)$ in this case.
Summing up, we have proved the following result.

\begin{theorem}
\label{FUN-COMP}
It holds $\FP_n(1) > \FP_n(0)$.
If $n \geq 2$, then the numbers $\FP_n(0),\ldots,\FP_n(n)$ form a chain
$
\FP_n(1) 
> \FP_n(0) 
\geq \FP_n(2) 
> \ldots 
> \FP_n(n)$
that starts with $\FP_n(1) = n (n-1)^{n-1}$ and ends 
with $\FP_n(n) = 1$.
If $n \geq 3$, then this chain is even strictly 
decreasing.
\end{theorem}

\noindent
For a description of the corresponding relationships for the list
$\FP^u_n(0),\ldots, \FP^u_n(n)$ obtained by Theorem \ref{THEEXIS-GEN}(b),
we first use Corollary \ref{LEM2GEN}(b) and $\binom{n}{1} = n$ and 
calculate
$
\FP^u_n(0) = n^n =  n n^{n-1} = \FP^u_n(1) ,
$
which means that on $X$ there exist exactly as many partial functions 
without a fixed point as partial functions with one fixed point. In Section \ref{Sec:PFunBij}
we will prove that these two sets are isomorphic in general, i.e., there is a bijection between
the set of partial functions on an arbitrary set $X$ without a fixed point and the set of partial functions 
on $X$ with one fixed point. In view of the remaining relationships, we prove again 
a recursive description, viz.
\begin{align*}
  \FP^u_n(k+1)
&= \binom{n}{k+1} n^{n-(k+1)} \\
&= \frac{n-k}{k+1} \binom{n}{k} n^{n-k-1} \\
&= \frac{n-k}{(k+1)n} \binom{n}{k} n^{n-k} \\
&= \frac{n-k}{(k+1)n} \FP^u_n(k),
\end{align*}
where $k \in \{0,\ldots,n-1\}$.
The calculation again uses Lemma \ref{BINOM}(a) and is also correct if 
$n = 1$.
For $k > 0$ this implies $\frac{n-k}{(k+1)n} < 1$ which in turn shows
$\FP^u_n(k) > \FP^u_n(k+1)$.
The following theorem summarizes these results.

\begin{theorem}
\label{PFUN-COMP}
It holds $\FP^u_n(0) = \FP^u_n(1)$.
If $n \geq 2$, then the numbers $\FP^u_n(0),\ldots,\FP^u_n(n)$ form a chain
$
\FP^u_n(0) 
= \FP^u_n(1) 
> \ldots 
> \FP^u_n(n)$
that starts with $\FP^u_n(0) = n^n$
and again ends with $\FP^u_n(n) = 1$.
\end{theorem}

\noindent
Next, we investigate how the members of the list
$\RP^t_n(0),\ldots,\RP^t_n(n)$ obtained by Theorem \ref{THEEXIS-GEN}(c)
are related to each other with respect to their sizes.
As a first relationship we have
$
\RP^t_n(0) = (2^{n-1} - 1)^n
        < (2^{n-1})^n
        = 2^{(n-1)n}
        = 2^{n^2 - n} 
        = \RP^t_n(n),
$
where the first step
uses $\binom{n}{0} = 1$.
With respect to the sublist $\RP^t_n(1),\ldots,\RP^t_n(n)$, we 
again calculate first a recursive description.
For this, we assume $n \geq 2$ and an arbitrary $k \in \{0,\ldots,n-1\}$.
Using Lemma \ref{BINOM}(a) in the second step, we get:
\begin{align*}
\RP^t_n(k+1) 
 &= \binom{n}{k+1} 
      2^{(k+1)(n-1)} (2^{n-1} - 1)^{n-(k+1)} \\
 &= \frac{n-k}{k+1} \binom{n}{k}
      2^{kn-k+n-1} (2^{n-1} - 1)^{n-k-1} \\
 &=  \frac{(n-k)2^{n-1}} {(k+1)(2^{n-1}-1)} 
     \binom{n}{k} 2^{kn-k} (2^{n-1} - 1)^{n-k} \\
 &=  \frac{(n-k)2^{n-1}} {(k+1)(2^{n-1}-1)} \RP^t_n(k) 
\end{align*}
Thus, we have
$\frac{\RP^t_n(k)}{\RP^t_n(k+1)}
= \frac{(k+1)(2^{n-1}-1)}{(n-k)2^{n-1}}$.
To determine when $\RP^t_n(k)$ is smaller or larger than $\RP^t_n(k+1)$
or both numbers are even equal,
we transform
$\frac{(k+1)(2^{n-1}-1)}{(n-k)2^{n-1}} \gtrless 1$ as follows,
where $\gtrless\, \in \{<,=,>\}$.
Note that due to the constraints $n \geq 2$ and $0 \leq k < n$,
all factors are positive.
\begin{align*}
\frac{(k+1)(2^{n-1}-1)}{(n-k)2^{n-1}} \gtrless 1
 & \iff (k+1)(2^{n-1}-1) \gtrless (n-k)2^{n-1} \\
 & \iff k 2^{n-1} - k + 2^{n-1} - 1 \gtrless n 2^{n-1} - k 2^{n-1} \\
 & \iff k (2^{n-1} + 2^{n-1} - 1) \gtrless n 2^{n-1} - 2^{n-1} + 1  \\
 & \iff k (2^n - 1) \gtrless (n-1) (2^{n-1} - \frac{1}{2}) + \frac{n + 1}{2} \\
 & \iff k (2^n - 1) \gtrless \frac{n-1}{2} (2^n - 1) + \frac{n + 1}{2} \\
 & \iff k \gtrless \frac{n-1}{2} + \frac{n+1}{2(2^n - 1)} 
\end{align*}
This simplifies to $k \gtrless 1$ for $n = 2$, hence
$\RP^t_2(0) < \RP^t_2(1) = \RP^t_2(2)$.
If $n>2$, then $\frac{n+1}{2(2^n - 1)} < \frac{1}{2}$,
which means that $\frac{n-1}{2} + \frac{n+1}{2(2^n - 1)}$ can never be an integer,
thus excluding $\RP^t_n(k) = \RP^t_n(k+1)$.
The remaining two cases are
\[
\RP^t_n(k) < \RP^t_n(k+1)
\iff k < \frac{n-1}{2} + \frac{n+1}{2(2^n - 1)}
\iff k \leq \left\lfloor \frac{n-1}{2} \right\rfloor
\]
and
\[
\RP^t_n(k) > \RP^t_n(k+1)
\iff k > \frac{n-1}{2} + \frac{n+1}{2(2^n - 1)}
\iff k > \left\lfloor \frac{n-1}{2} \right\rfloor
\]
(since $\frac{n+1}{2(2^n - 1)} < \frac{1}{2}$ and the decimal place of $\frac{n-1}{2}$ is either $0$ or $5$, the $\frac{n+1}{2(2^n - 1)}$ part can be 
ignored when taking the floor).
See Figure~\ref{fig:total-relations-fixed-points} for a few values.
\begin{figure}[ht]
\[
\begin{array}{r|c|rrrrrr}
\sum\limits_k & n \setminus k  &      0 &       1 &       2 &       3 &       4 &       5 \\\hline
       1 & 1 &      0 &       1 &&&&\\
       9 & 2 &      1 &       4 &       4 &&&\\
     343 & 3 &     27 &     108 &     144 &      64 &&\\
   50625 & 4 &   2401 &   10976 &   18816 &   14336 &    4096 &\\
28629151 & 5 & 759375 & 4050000 & 8640000 & 9216000 & 4915200 & 1048576 \\
\end{array}
\]
\caption{Number of total relations on an $n$-set with $k$ reflexive points}
\label{fig:total-relations-fixed-points}
\end{figure}
The second column contains the numbers $n$, where $1 \leq n \leq 5$.
The row for $n$, from the third element to the end,
then contains the numbers $\RP^t_n(0)$ to $\RP^t_n(n)$ 
and the first element of this row is the sum of these numbers.
What we have shown can be summarized as follows.

\begin{theorem}
\label{TOTREL-COMP}
It holds $\RP^t_n(0) < \RP^t_n(n)$.
If $n=2$, then $\RP^t_n(0) < \RP^t_n(1) = \RP^t_n(2)$.
If $n>2$, then the sequence
$\RP^t_n(0), \ldots < \RP^t_n(n)$
strictly increases till $\lfloor \frac{n-1}{2} \rfloor + 1$
and then strictly decreases.
The ascending chain starts with
$\RP^t_n(0) = (2^{n-1} - 1)^n$ and the descending chain
ends with
$\RP^t_n(n) = 2^{n^2 - n}$.
\end{theorem}

\noindent
For an odd $n$ the ascending chain has one
element more than the descending chain.
E.g., for $n=3$ we have 
$27 < 108 < 144$ for $\RP^t_3(0), \RP^t_3(1), \RP^t_3(2)$ and 
$144 > 64$ for $\RP^t_3(2), \RP^t_3(3)$.
See again Figure~\ref{fig:total-relations-fixed-points}.

To describe the relationships between the numbers of
the list $\RP_n(0),\ldots,\RP_n(n)$
obtained by Theorem \ref{THEEXIS-GEN}(d),
we notice that from their definition it follows that
they are related to each other with respect to their sizes 
exactly as the binomial coefficients of the $n$-th row of
Pascal's triangle are, because $2^{n^2-n}$ does not depend on $k$.
This shows:

\begin{theorem}
\label{REL-COMP}
For all $k \in \{0,\ldots,n\}$ it holds
$\RP_n(k) = \RP_n(n-k)$.
If $n \geq 2$ is even, then the numbers $\RP_n(0),\ldots,\RP_n(n)$ 
form the chains
$\RP_n(0) < \ldots < \RP_n(\frac{n}{2})$
and 
$\RP_n(\frac{n}{2}) > \ldots > \RP_n(n)$, and 
if $n \geq 2$ is odd, then they form the chains
$\RP_n(0) < \ldots < \RP_n(\frac{n-1}{2})$
and 
$\RP_n(\frac{n+1}{2}) > \ldots > \RP_n(n)$, where
$\RP_n(\frac{n-1}{2}) = \RP_n(\frac{n+1}{2})$.
Both ascending chains start with
$\RP_n(0) = 2^{n^2 - n}$
and both descending chains end with
$\RP_n(n) = 2^{n^2 - n}$.
\end{theorem}

\noindent
Using Lemma \ref{BINOM}(a), also for the numbers $\RP_n(0),\ldots,\RP_n(n)$ 
a recursive description can be shown, viz.\ that
$
\RP_n(k+1) 
= \frac{n-k}{k+1} \RP_n(k),
$
for all $k \in \{0,\ldots,n-1\}$.

Finally, in the following we compare the members of the list
$\FP^p_n(0) ,\ldots, \FP^p_n(n)$ 
obtained by Theorem \ref{THEPERM-GEN}
with respect to their sizes.
For that, we assume an arbitrary $k \in \{0,\ldots,n-1\}$.
Then we have:
\begin{align*}
\FP^p_n(k) > \FP^p_n(k+1)
  & \iff \binom{n}{k} \,\cdot\, !(n-k) > \binom{n}{k+1} \,\cdot\, !(n-(k+1))
    & \mbox{def. } \FP^p_n(i) \\
  & \iff \binom{n}{k} \,\cdot\, !(n-k) > 
         \frac{n-k}{k+1} \binom{n}{k} \,\cdot\, !(n-k-1)
    & \mbox{Lemma } \ref{BINOM}\mbox{(a)} \\
  & \iff (k+1) \,\cdot\, !(n-k) > (n-k) \,\cdot\, !(n-k-1)
    \\
  & \iff k \,\cdot\, !(n-k)~+~!(n-k)~>~!(n-k) - (-1)^{n-k}
    & \mbox{Lemma } \ref{SUBFAC}\mbox{(a)} \\
  & \iff k \,\cdot\, !(n-k) + (-1)^{n-k} > 0 
\end{align*}
Next, we determine when $k$ $\cdot$ $!(n-k)$ $+$ $(-1)^{n-k} > 0$ holds.
We consider two cases.
To help understand them, we list the values of $!n$ for 
$0 \leq n \leq 10$ in the table of
Figure \ref{fig:SUBFACF0-10}
(see also~\cite{Stanley2012,Wiki-derangement}, for example):
\begin{figure}[ht]
\[
\begin{array}{ccccccccccc}
0 & 1 & 2 & 3 & 4 & 5 & 6 & 7 & 8 & 9 & 10 \\\hline
1 & 0 & 1 & 2 & 9 & 44 & 265 & 1854 & 14833 & 133496 & 1334961 \\
\end{array}
\]
\caption{Subfactorials $!n$ for $0 \leq n \leq 10$}
\label{fig:SUBFACF0-10}
\end{figure}
As first case we assume $n - k$ to be even.
Then $k$ $\cdot$ $!(n-k)$ $+$ $(-1)^{n-k} > 0$ holds by $!(n-k) \geq 0$ and 
$(-1)^{n-k} = 1 > 0$.
The second case is that $n - k$ is odd.
In this case $k~\cdot~!(n-k) + (-1)^{n-k} > 0$ is equivalent to
$k~\cdot~!(n-k) > 1$.
The latter property fails iff $k = 0$ or $n - k = 1$ since in those cases one of the factors on the left-hand side is equal to $0$ rather than positive.
Combining the two cases, we conclude that
$\FP^p_n(k) > \FP^p_n(k+1)$ holds
for the selected $k \in \{0,\ldots,n-1\}$ 
except when $n$ is odd and $k = 0$ or when $k = n - 1$.
If $n$ is odd and $k = 0$, then
$\FP^p_n(0) = \FP^p_n(1) - 1 < \FP^p_n(1)$,
and if $k = n - 1$, then
$\FP^p_n(n-1) = 0 < 1 = \FP^p_n(n)$.
All that leads to the following result.

\begin{theorem}
\label{PERM-COMP}
It holds $\FP^p_n(n-1) < \FP^p_n(n)$.
If $n \geq 2$ is even, then the numbers $\FP^p_n(0),\ldots,\FP^p_n(n-1)$ 
form a chain $\FP^p_n(0) > \ldots > \FP^p_n(n-1)$,
and if $n \geq 2$ is odd, then
$\FP^p_n(0) < \FP^p_n(1)$ and the
numbers $\FP^p_n(1),\ldots,\FP^p_n(n-1)$ 
form a chain $\FP^p_n(1) > \ldots > \FP^p_n(n-1)$.
The first chain starts with $\FP^p_n(0)$ $=$ $!n$ and 
the second one starts with 
$\FP^p_n(1)$ $=$ $n$ $\cdot$ $!(n-1)$.
Both chains end with
$\FP^p_n(n-1) = 0$.
\end{theorem}

\begin{figure}[ht]
\[
\begin{array}{r|rrrrrrrrrrr}
n \setminus k  &       0 &       1 &      2 &      3 &     4 &     5 &    6 &   7 &  8 & 9 & 10 \\
\hline
 1 &       0 &       1 \\
 2 &       1 &       0 &      1 \\
 3 &       2 &       3 &      0 &      1 \\
 4 &       9 &       8 &      6 &      0 &     1 \\
 5 &      44 &      45 &     20 &     10 &     0 &     1 \\
 6 &     265 &     264 &    135 &     40 &    15 &     0 &    1 \\
 7 &    1854 &    1855 &    924 &    315 &    70 &    21 &    0 &   1 \\
 8 &   14833 &   14832 &   7420 &   2464 &   630 &   112 &   28 &   0 &  1 \\
 9 &  133496 &  133497 &  66744 &  22260 &  5544 &  1134 &  168 &  36 &  0 & 1 \\
10 & 1334961 & 1334960 & 667485 & 222480 & 55650 & 11088 & 1890 & 240 & 45 & 0 &  1 \\
\end{array}
\]
\caption{Number of permutations on an $n$-set with $k$ fixed points}
\label{fig:permutations-fixed-points}
\end{figure}

\noindent
To give an impression on the size of the numbers and their
relationships, we present the values of
$\FP^p_n(0)$ to $\FP^p_n(n)$ for $1 \leq n \leq 10$ in
the table of Figure \ref{fig:permutations-fixed-points}.


\section{Involutions and Proper Involutions}
\label{WITHOUTFP}

A function $f : X \to X$ is an involution on the set $X$ if 
$f(f(x)) = x$, for all $x \in X$.
If $f$ is taken as a relation on $X$, this is equivalent to 
$f \Comp f = \I_X$.
An involution is said to be proper if it has no fixed point.
Involutions are specific (viz. self-inverse) permutations.
The number $\INV_n$ of involutions on a set with $n$ elements 
satisfies the recursion $\INV_0 = \INV_1 = 1$ and 
$\INV_n = \INV_{n-1} + (n-1)\INV_{n-2}$,
for all $n \in \NAT_{\geq 2}$
(see \cite{Knuth,WeissteinPI,WikiInvolution,WikiTelephone}).
The first 28 members of the sequence 
$(\INV_n)_{n \geq 0}$ constitute the sequence A000085 
in the OEIS data base (see \cite{A000085}).

An involution matches some pairs of elements and leaves the other 
elements untouched. 
In other words, in the directed graph representation of an involution
all elements are part of a cycle, some of length~$2$ and the 
others of length $1$, i.e., of a loop.
Exactly the latter are the fixed points.

First, we count the number of involutions on a set with $n$ elements 
that have $k$ fixed points.
To this end, for all $n \in \NAT$ and $k \in \{0, \ldots n\}$ 
we define the number $\INV_n(k)$ as follows:
\begin{equation}
\label{eq:J_n(k)}
\INV_n(k) := \left\{\begin{array}{cl}
                 0 & \mbox{if $n - k$ is odd} \\
                 \frac{n!}
                      {2^{\frac{n-k}{2}} (\frac{n-k}{2}) ! k!} & \mbox{otherwise}
                 \end{array}\right.
\end{equation}
Notice, that in the second case the power $2^{\frac{n-k}{2}}$
is a natural number and the factorial $(\frac{n-k}{2})!$ is defined,
since $n-k$ is even and $k \leq n$.
Based upon the above definition, we get the following result.

\begin{theorem}
\label{THENUMBPINV}
Assume $n \in \NAT$, $X$ be a finite set with $n = |X|$ and
$k \in \{0, \ldots n\}$. 
Then the number of involutions on $X$ with $k$ fixed points is $\INV_n(k)$.
\end{theorem}
\begin{proof}
It is known that
$\frac{n!}{2^m m! (n - 2m)!}$
is the number of involutions on a set with $n$ elements and 
$m$ matched pairs (see e.g.,~\cite{WikiTelephone}).
For each such involution $n - 2m$ is the number of fixed points,
hence $k = n - 2m$, hence $m = \frac{n - k}{2}$.
This shows that there is no involution if $n - k$ is odd.
For even $n - k$, the result follows by replacing $m$ by 
$\frac{n - k}{2}$ in $\frac{n!}{2^m m! (n - 2m)!}$, thus 
obtaining the definition of $\INV_n(k)$. 
\end{proof}

\noindent
An immediate consequence of Theorem \ref{THENUMBPINV} is
that the number of proper involutions on a finite set with $n$ 
elements is $\INV_n(0)$, where $\INV_n(0) = 0$, if $n$ is odd, and 
$\INV_n(0) = \frac{n!}
               {2^{\frac{n}{2}} 
                \frac{n}{2} !}$, 
if $n$ is even.

\begin{figure}
\begin{center}
\rotatebox{90}{
{
\scriptsize
$
\begin{array}{l}
\begin{array}{r||r||r|r|r|r|r|r|r|r|r|r|r|r|r|r|r|}
\sum\limits_k & n \setminus k & 0 & 1 & 2 & 3 & 4 & 5 & 6 & 7 & 8 & 9 & 10 & 11 & 12 & 13 & 14 \\
\hline\hline
1   & 0 & 1 &&&&&&&&&&&&&& \\
1   & 1 &   0 & 1 &&&&&&&&&&&&& \\
2   & 2 &   1 &   0 &   1 &&&&&&&&&&&& \\
4   & 3 &   0 &   3 &   0 &   1 &&&&&&&&&&& \\
10  & 4 &   3 &   0 &   6 &   0 &   1 &&&&&&&&&& \\
26  & 5 &   0 &  15 &   0 &  10 &   0 &  1 &&&&&&&&& \\
76  & 6 &  15 &   0 &  45 &   0 &  15 &  0 &  1 &&&&&&&& \\
232 & 7 &   0 & 105 &   0 & 105 &   0 & 21 &  0 & 1 &&&&&&& \\
764 & 8 & 105 &   0 & 420 &   0 & 210 &  0 & 28 & 0 & 1 &&&&&& \\
2620 & 9 & 0 & 945 & 0 & 1260 & 0 & 378 & 0 & 36 & 0 & 1 &&&&& \\
9496 & 10 & 945 & 0 & 4725 & 0 & 3150 & 0 & 630 & 0 & 45 & 0 & 1 &&&& \\
35696 & 11 & 0 & 10395 & 0 & 17325 & 0 & 6930 & 0 & 990 & 0 & 55 & 0 & 1 &&& \\
140152 & 12 & 10395 & 0 & 62370 & 0 & 51975 & 0 & 13860 & 0 & 1485 & 0 & 66 & 0 & 1 && \\
568504 & 13 & 0 & 135135 & 0 & 270270 & 0 & 135135 & 0 & 25740 & 0 & 2145 & 0 & 78 & 0 & 1 & \\
2390480 & 14 & 135135 & 0 & 945945 & 0 & 945945 & 0 & 315315 & 0 & 45045 & 0 & 3003 & 0 & 91 & 0 & 1 \\
\end{array}
\\\\\\
\begin{array}{r||r||r|r|r|r|r|r|r|r|r|}
n & 0  & 2 & 4 & 6 & 8 & 10 & 12 & 14 & 16 & 18 \\
\hline\hline
\mbox{\# proper (column 0)} & 1 & 1 & 3 & 15 & 105 & 945 & 10395 & 135135 & 2027025 & 34459425 \\
\mbox{\# involutions (column $\sum\limits_k$)} & 1 & 2 & 10 & 76 & 764 & 9496 & 140152 & 2390480 & 46206736 & 997313824 \\
\mbox{Ratio} & 100.00\% & 50.00\% & 30.00\% & 19.74\% & 13.743\% & 9.952\% & 10395\% & 5.653\% & 4.387\% & 3.455\% \\
\end{array}
\end{array}
$
}
}
\end{center}
\vspace{-5mm}
\caption{$\INV_n(k)$ with $\sum\limits_k \INV_n(k)$ and ratios}
\label{fig:J_n(k)-ratios}
\end{figure}

In Figure~\ref{fig:J_n(k)-ratios} the upper table gives the number 
of involutions with $k$ fixed points and $0 \leq n \leq 14$. 
All values of $k$ label the columns and those of $n$ (column 2) label the rows.
The first column contains the sum $\sum\limits_{k=0}^n \INV_n(k)$; this 
is the total number $\INV_n$ of involutions on a set with $n$ elements. 
As $\INV_n(0)$ is the number of proper involutions,
the column labeled with $k = 0$ contains the number of these functions.
For $0 \leq n \leq 18$ and a set with $n$ elements, the lower table 
gives the number of proper involutions $\INV_n(0)$, the total number 
of involutions and the ratio of $\INV_n(0)$ to the total number of
involutions.
These numbers are given only for even $n$, since $\INV_n(0)$ and 
the ratio  are~$0$ for odd $n$. The numbers in the line ``\# proper''
are the first ten of the sequence A001147
in the OEIS data base (see \cite{A001147}).
The ratio seems to decrease to~$0$ as $n$ tends to infinity.
This is proved in the following. 
Doing so, we also compare the numbers $\INV_n(k+2)$ and $\INV_n(k)$, 
along with results on various limiting values.
We start with the following auxiliary result that gives a simple
description of the ratio of $\INV_n(k)$ to $\INV_n(k+2)$ in case
that $n-k$ is even.

\begin{lemma}
\label{E-simplification}
Assume $n \in \NAT$ and a finite set $X$ with $n = |X|$.
For all $k \in \{0, \ldots n-2\}$ define
$E_n(k) := \frac{\INV_n(k)}{\INV_n(k+2)}$.
If $n - k$ is even, then $E_n(k) = \frac{(k+1)(k + 2)}{n - k}$.
\end{lemma}
\begin{proof}
Assume that $n - k$ is even. 
Then the claim is shown by
$$
E_n(k)
= \frac{\INV_n(k)}{\INV_n(k+2)}
= \frac{2^{\frac{n-k}{2} - 1} (\frac{n-k}{2} - 1)! (k + 2)!}{
            2^{\frac{n-k}{2}} (\frac{n-k}{2})! k!}
= \frac{(k+1)(k + 2)}{n - k},
$$
where the definitions of $\INV_n(k)$ and $\INV_n(k+2)$ and some
simple algebraic calculations are used.
\end{proof}

\noindent
With the help of this lemma we now show the next result.
Especially, it gives a simple
description of the relationship
between $\INV_n(k)$ and $\INV_n(k+2)$ in case that $n - k$ is even.
So, the following theorem is the counterpart to 
the Theorems \ref{FUN-COMP} to \ref{PERM-COMP}.

\begin{theorem}
\label{J_n(k)-increase-decrease}
Let $n \in \NAT_{\geq 1}$, a  finite set $X$ with $n = |X|$
and $k \in \{0, \ldots n-2\}$ be given.
If $n-k$ is odd, then $\INV_n(k) = 0 = \INV_n(k+2)$, and
if $n-k$ is even, then 
\begin{enumerate}
\item[(a)] $\INV_n(k) < \INV_n(k+2)$ iff $k < \sqrt{n+2} - 2$,
\item[(b)] $\INV_n(k) = \INV_n(k+2)$ iff $k = \sqrt{n+2} - 2$ and
\item[(c)] $\INV_n(k) > \INV_n(k+2)$ iff $k > \sqrt{n+2} - 2$.
\end{enumerate}
\end{theorem}
\begin{proof}
If $n - k$ is odd, then $n - (k+2)$ is odd, too, hence
$\INV_n(k) = 0 = \INV_n(k+2)$.
Suppose now that $n - k$ is even.
Using again $\gtrless$ $\in$ $\{<,=,>\}$, 
we start the proof of the equivalences (a), (b) and (c) with the 
following calculation, where
Lemma \ref{E-simplification} is used in the second step:
\begin{align*}
\INV_n(k) \gtrless \INV_n(k+2)
& \iff \frac{\INV_n(k)}{\INV_n(k+2)} \gtrless 1 \\
& \iff \frac{(k+1)(k + 2)}{n - k} \gtrless 1 \\
& \iff (k + 1)(k + 2) \gtrless n - k \\
& \iff k^2 + 4k + 2 - n \gtrless 0 
\end{align*}
As a consequence, it remains to show that 
$k^2 + 4k + 2 - n \gtrless 0$ iff $k \gtrless \sqrt{n+2} - 2$.
For this, we consider the quadratic equation 
$x^2 + 4x + 2 - n = 0$.
It has two solutions, viz. $x_1 = -2 + \sqrt{n+2}$ and
$x_2 = -2 - \sqrt{n+2}$.
Hence, as $k$ is a natural number, 
$k^2 + 4k + 2 - n < 0$ iff $k < \sqrt{n+2} - 2$ and
$k^2 + 4k + 2 - n = 0$ iff $k = \sqrt{n+2} - 2$ and
$k^2 + 4k + 2 - n > 0$ iff $k > \sqrt{n+2} - 2$.
\end{proof}

\noindent
How the numbers of the list $\INV_n(0),\ldots,\INV_n(n)$ for a given
$n$ are related
to each other with respect to their sizes can be summarized as 
follows:
If $n$ is even, then $\INV_n(k) = 0$ for odd $k$.
The numbers of the sublist $\INV_n(0),\INV_n(2),\ldots,\INV_n(n)$ 
for even $k$ are increasing until $k$ exceeds $\sqrt{n+2} - 2$.
Then they are decreasing.
If $n$ is odd, then $\INV_n(k) = 0$ for even $k$.
The numbers of the sublist $\INV_n(1),\INV_n(3),\ldots,\INV_n(n)$ 
for odd $k$ are again increasing until $k$ exceeds $\sqrt{n+2} - 2$
and then again decreasing.
If $\sqrt{n+2} - 2$ is a natural number, then the last element
of the increasing part equals the first element of the decreasing 
part.
This happens, for instance, if $n = 14$, as $\sqrt{14+2} - 2 = 2$. 
{From} $0 < 2$ and Theorem \ref{J_n(k)-increase-decrease}(a)
we get $\INV_{14}(0) < \INV_{14}(2)$.
Then part (b) of the theorem yields
$\INV_{14}(2) = \INV_{14}(4)$. 
Finally, part (c) leads to the chain
$\INV_{14}(4) > \INV_{14}(6) > \ldots > \INV_{14}(14)$,
since $14 > \ldots > 6 > 4 > 2$.
Compare with Figure~\ref{fig:J_n(k)-ratios}.

Next, we consider again the ratio of $\INV_n(k)$ to $\INV_n(k+2)$,
i.e., the number $E_n(k) := \frac{\INV_n(k)}{\INV_n(k+2)}$ introduced in 
Lemma \ref{E-simplification}, in case that $n-k$ is even.
For a fixed $k$ and $n$ tending to infinity, we are interested in the 
limiting value of the subsequence of
$(E_n(k))_{n \geq 2}$
that contains 
those numbers only for which $n-k$ is even.
In the following lemma we use the notation 
$\lim\limits_{n - k\textrm{ even}, n \to \infty} E_n(k)$ for this specific
limiting value.

\begin{lemma}
\label{limit-E}
Let $k \in \NAT$ be fixed.
Then
$\lim\limits_{n - k\textrm{ even}, n \to \infty} E_n(k) = 0$.
\end{lemma}
\begin{proof}
The result is a direct consequence of Lemma \ref{E-simplification},
since for $k$ fixed the numerator of $\frac{(k+1)(k + 2)}{n - k}$
is a constant and the denominator of this fraction tends to infinity 
if $n$ tends to infinity.
\end{proof}

\noindent
It may seem strange that the proof of Lemma \ref{limit-E} does not 
distinguish the zone where the numbers $\INV_n(k)$, with $n-k$ even, 
increase from the zone where they strictly decrease. 
The reason is that for a fixed $k$ the point where the numbers $\INV_n(k)$,
with $n-k$ even, start
to strictly
decrease, i.e., $\sqrt{n+2} - 2$ by Theorem \ref{J_n(k)-increase-decrease}, 
becomes larger than the fixed $k$ for large enough $n$.

Assume $X$ to be a finite set with
$n \geq 0$ elements and that functions 
on $X$ are chosen uniformly at random.
Then $\frac{\INV_n(0)}{\INV_n}$ is the probability that a randomly chosen 
involution on $X$ is proper,
$\frac{\INV_n - \INV_n(0)}{\INV_n}$ is the probability that a randomly chosen 
involution on $X$ has a fixed point and,
with $\FP^p_n(k) := \binom{n}{k} \cdot~!(n-k)$ defined as in 
Section \ref{WITHFP}, $\frac{\INV_n(k)}{\FP^p_n(k)}$ is the probability 
that a randomly chosen permutation on $X$ that has $k$ fixed points 
is an involution (with $k$ fixed points).
In the next theorem, we present the limiting values of these probabilities if
$n$ tends to infinity.

\begin{theorem}
\label{th:limits-involutions}
We have:
\begin{enumerate}[(a)]
\item $\lim\limits_{n \to \infty} \frac{\INV_n(0)}{\INV_n} = 0$.

\item $\lim\limits_{n \to \infty} \frac{\INV_n - \INV_n(0)}{\INV_n} = 1$.
\item $\lim\limits_{n \to \infty} \frac{\INV_n(k)}{\FP^p_n(k)} = 0$,
      for all $k \in \NAT$.
\end{enumerate}
\end{theorem}
\begin{proof}
(a) By~\eqref{eq:J_n(k)}, $\INV_n(0) = 0$ if $n$ is odd.
As a consequence, for the limiting value we are interested in only 
those numbers of the sequence $(\frac{\INV_n(0)}{\INV_n})_{n\geq 0}$ 
have to be considered that have an even index.
If we use for this specific limiting value a notation similar 
to that of Lemma \ref{limit-E}, this means
$
\lim\limits_{n \to \infty} \frac{\INV_n(0)}{\INV_n} = 
\lim\limits_{n \textrm{ even},n \to \infty} \frac{\INV_{n}(0)}{\INV_{n}}
$
and, hence, we only have to prove 
$\lim\limits_{n \textrm{ even},n \to \infty} \frac{\INV_{n}(0)}{\INV_{n}} = 0$.
The number $\INV_{n}$ is given as
$\INV_{n} = \sum\limits_{k=0}^{n} \INV_{n}(k)$. 
Now, the claim follows from the calculation
\[
\lim_{n \textrm{ even},n \to \infty} \frac{\INV_{n}(0)}{\INV_{n}}
= \lim_{n \textrm{ even},n \to \infty} \frac{\INV_{n}(0)}{\sum\limits_{k=0}^{n} \INV_{n}(k)}
\leq \lim_{n \textrm{ even},n \to \infty} \frac{\INV_{n}(0)}{\INV_{n}(2)}
= \lim_{n \textrm{ even},n \to \infty} E_{n}(0)
= 0,
\]
where the definition of $E_n(0)$ and Lemma~\ref{limit-E} are 
used in the last two steps.

(b) By part (a) we get
$$
\lim_{n \to \infty} \frac{\INV_n - \INV_n(0)}{\INV_n} 
= \lim_{n \to \infty} 1 - \frac{\INV_n(0)}{\INV_n} 
= 1 - \lim_{n \to \infty} \frac{\INV_n(0)}{\INV_n} 
= 1 - 0
= 1.
$$

(c) Assume an arbitrary $k \in \NAT$.
For all $n \in \NAT$ such that $n - k$ is odd we have $\INV_n(k) = 0$.
To show $\lim\limits_{n \to \infty} \frac{\INV_n(k)}{\FP^p_n(k)} = 0$, therefore
it suffices to prove
$
\lim\limits_{n-k \textrm{ even},n \to \infty} \frac{\INV_n(k)}{\FP^p_n(k)} = 0, 
$
where again a notation similar to that of Lemma \ref{limit-E} is used.
Suppose $n - k \geq 4$. 
This is not a restriction, since we want the
limiting value when $n$ tends 
to infinity for a fixed $k$.
Then we calculate as follows:
\begin{align*}
\frac{\INV_n(k)}{\FP^p_n(k)}
& = \frac{~~ \frac{n!}{2^{\frac{n-k}{2}} (\frac{n-k}{2}) ! k!}~~}{ \binom{n}{k} \cdot~!(n-k)}
   && \mbox{definition of $\INV_n(k)$ and $\FP^p_n(k)$} \\
& = \frac{n!}{2^{\frac{n-k}{2}} (\frac{n-k}{2})! k! \binom{n}{k} \lfloor \frac{(n - k)! + 1}{\rm e}\rfloor}
    && \mbox{Lemma \ref{SUBFAC}(b)} \\
& \leq \frac{n!}{2^{\frac{n-k}{2}} (\frac{n-k}{2})! k! \binom{n}{k} \lfloor \frac{(n - k)!}{\rm 4}\rfloor}
    && \mbox{${\rm e} \leq 4$ and dropping $+ 1$} \\
& = \frac{n!}{2^{\frac{n-k}{2}} (\frac{n-k}{2})! k! \binom{n}{k} \frac{(n - k)!}{4}}
    && \mbox{$ \frac{(n - k)!}{4} \in \NAT$ since $n - k \geq 4$} \\
& \leq \frac{4\,n!}{2^{\frac{n-k}{2}} k! \binom{n}{k}(n - k)!} \\
& = \frac{4 \binom{n}{k}}{2^{\frac{n-k}{2}}\binom{n}{k}}
    && \mbox{definition of $\binom{n}{k}$} \\
& = \frac{4}{2^{\frac{n-k}{2}}}
\end{align*}
This implies
$
\lim\limits_{n-k \textrm{ even},n \to \infty} \frac{\INV_n(k)}{\FP^p_n(k)} 
= \lim\limits_{n-k \textrm{ even},n \to \infty} \frac{4}{2^{\frac{n-k}{2}}}
= 0
$.
\end{proof}


\section{Idempotent and Transitive Partial Functions}
\label{IDEMP}

A function $f : X \to X$ is idempotent if $f(f(x)) = f(x)$, for 
all $x \in X$.
Using relational composition this means that $f \Comp f = f$.
When $f : X \to X$ is a partial function with $f \Comp f = f$, it is an
idempotent partial function, and when only $f \Comp f \subseteq f$ holds,
it is a transitive partial function.
Clearly, if $f$ is idempotent, then $f(x)$ is a fixed point of 
$f$, for all $x \in X$, and if $f$ is transitive and $x \in X$, 
then either $f(x)$ is a fixed point of $f$ or $f(x) \notin \dom{f}$.
A function is idempotent iff it is transitive.
So, a discrimination is only interesting in case of partial functions.
We start the investigations of these classes of relations with the
following result.
In the statement of the theorem, an out-domain element of a
partial function $f : X \to X$ is an element of $X$ that 
is not contained in $\dom{f}$.

\begin{theorem}
\label{th:transitive-functions}
Assume $n \in \NAT_{\geq 1}$, $k, l \in \{0,\ldots,n\}$ 
and a finite set $X$ with $n = |X|$.
Then the number of transitive partial functions $f : X \to X$ 
with $k$ fixed points and $l$ out-domain elements is
$
\binom{n}{k} \binom{n - k}{l} (k+l)^{n-(k+l)}
$.
\end{theorem}

\begin{proof}
The result is obtained by first choosing $k$ fixed points from $n$ 
elements and then choosing $l$ elements from the remaining $n - k$ 
elements. 
The value of $f$ on the $k$ fixed points is determined, so there 
is nothing to choose there. 
For the other $n-(k+l)$ elements $x$, where $f(x)$ is defined, 
there are $k+l$ possibilities for $f(x)$, since $f(x)$ 
is either a fixed point or an element not in $\dom{f}$.
\end{proof}

\noindent
Another possible proof consists in first choosing the $l$ elements 
not in $\dom{f}$ and then choosing the $k$ fixed points from 
the remaining $n-l$ elements.
We obtain
$\binom{n}{l} \binom{n - l}{k} (k+l)^{n-(k+l)}$
and this gives the same result, since
$
\binom{n}{k} \binom{n - k}{l}
= \frac{n!}{k!\, l!\, (n-(k+l))!}
= \binom{n}{l} \binom{n - l}{k}
$.
The middle expression displays the symmetry between $k$ and $l$.
In the following corollary we present four special cases of 
Theorem~\ref{th:transitive-functions}, where we assume $n$, $k$
and $l$ as in the theorem.

\begin{corollary}
\label{cor:transitive-functions}
\begin{enumerate}[(a)]
\item
\label{spec-idempotent}
The number of idempotent partial functions with $k$ fixed points and 
$l$ out-domain elements is 
$\binom{n}{k} \binom{n - k}{l} k^{n-(k+l)}$.

\item
The number of transitive partial functions with $k$ fixed points is
$\binom{n}{k} \sum\limits_{l=0}^{n-k} \binom{n - k}{l}(k+l)^{n-(k+l)}$.

\item
\label{spec-total}
The number of idempotent functions with $k$ fixed points is
$\binom{n}{k}k^{n-k}$.

\item
\label{spec-no-fp}
The number of transitive partial functions that have no fixed 
point and $l$ out-domain elements is $\binom{n}{l} l^{n-l}$.
\end{enumerate}
\end{corollary}

\begin{proof}
(a)
This number is obtained by a slight variation of the 
proof of Theorem \ref{th:transitive-functions}.
If $f$ is an idempotent partial function, then $f(x)$ 
must be a fixed point when it is defined.
Hence, we have to take $l = 0$ in the expression $k+l$
in the base of the exponentation $(k+l)^{n-(k+l)}$, 

(b)
This is a direct consequence of Theorem \ref{th:transitive-functions} 
and distributivity.

(c)
An idempotent function $f$ is total, i.e., no 
out-domain elements exist.
As a consequence, the result follows by taking $l = 0$ in 
Theorem \ref{th:transitive-functions}.

(d)
To have no fixed point means that we have
to take $k = 0$ in Theorem \ref{th:transitive-functions} to
get the result.
\end{proof}

\noindent
{From} part (\ref{spec-idempotent}) we get
$\binom{n}{k} \sum\limits_{l=0}^{n-k} \binom{n - k}{l} k^{n-(k+l)}$
as the number of idempotent partial functions on an $n$-set
that have $k$ fixed points.
An immediate (and obvious) consequence of part (c) is that
(due to the assumption $n \geq 1$) there is no idempotent 
function with no fixed point on a non-empty set.
When we also allow $n = 0$, then the unique function on the 
$0$-set $\emptyset$ is idempotent, since it satisfies $f \Comp f = f$.

There are two tables in Figure~\ref{fig:transitive-idempotent-functions}.
The left one displays numbers for transitive partial functions 
on an $11$-set.
An entry in line $l$ and column $k$ gives the number of such functions 
with $l$ out-domain elements and $k$ fixed points. 
The symmetry between $l$ and $k$ is very visible in this table.
The table on the right contains numbers for idempotent functions.
An entry in line $n$ and column $k$ is the number of such 
functions on an $n$-set with $k$ fixed points. 
Note that the line $n = 11$ is the same as the line $l = 0$ in 
the left table (except for $k = 0$).

\begin{figure}
\begin{center}
\rotatebox{90}{
{
\scriptsize
$
\begin{array}{l}
\begin{array}{r|rrrrrrrrrrrr}
l \setminus k &        0 &        1 &         2 &         3 &         4 &        5 &        6 &       7 &      8 &    9 &  10 & 11 \\\hline
 0 &        0 &       11 &     28160 &   1082565 &   5406720 &  7218750 &  3592512 &  792330 &  84480 & 4455 & 110 &  1 \\
 1 &       11 &    56320 &   3247695 &  21626880 &  36093750 & 21555072 &  5546310 &  675840 &  40095 & 1100 &  11 & \\
 2 &    28160 &  3247695 &  32440320 &  72187500 &  53887680 & 16638930 &  2365440 &  160380 &   4950 &   55 && \\
 3 &  1082565 & 21626880 &  72187500 &  71850240 &  27731550 &  4730880 &   374220 &   13200 &    165 &&& \\
 4 &  5406720 & 36093750 &  53887680 &  27731550 &   5913600 &   561330 &    23100 &     330 &&& \\
 5 &  7218750 & 21555072 &  16638930 &   4730880 &    561330 &    27720 &      462 &&&& \\
 6 &  3592512 &  5546310 &   2365440 &    374220 &     23100 &      462 &&&&& \\
 7 &   792330 &   675840 &    160380 &     13200 &       330 &&&&&& \\
 8 &    84480 &    40095 &      4950 &       165 &&&&&&& \\
 9 &     4455 &     1100 &        55 &&&&&&&& \\
10 &      110 &       11 &&&&&&&&& \\
11 &        1 &&&&&&&&&& \\\hline
\sum\limits_l & 18210094 & 88843084 & 180961110 & 199597200 & 129618060 & 50733144 & 11902044 & 1642080 & 129690 & 5610 & 121 &  1 \\
\end{array}
\\[40mm]
\begin{array}{r|r|rrrrrrrrrrrr}
\sum\limits_k & n \setminus k& 1 & 2 & 3 & 4 & 5 & 6 & 7 & 8 & 9 & 10 & 11 & 12 \\\hline
1 &  1 &  1 &&&&&&&&&&& \\
3 &  2 &  2 & 1 &&&&&&&&&& \\
10 &  3 &  3 & 6 & 1 &&&&&&&&& \\
41 &  4 &  4 & 24 & 12 & 1 &&&&&&&& \\
196 &  5 &  5 & 80 & 90 & 20 & 1 &&&&&&& \\
1057 &  6 &  6 & 240 & 540 & 240 & 30 & 1 &&&&&& \\
6322 &  7 &  7 & 672 & 2835 & 2240 & 525 & 42 & 1 &&&&& \\
41393 &  8 &  8 & 1792 & 13608 & 17920 & 7000 & 1008 & 56 & 1 &&&& \\
293608 &  9 &  9 & 4608 & 61236 & 129024 & 78750 & 18144 & 1764 & 72 & 1 &&& \\
2237921 & 10 & 10 & 11520 & 262440 & 860160 & 787500 & 272160 & 41160 & 2880 & 90 & 1 && \\
18210094 & 11 & 11 & 28160 & 1082565 & 5406720 & 7218750 & 3592512 & 792330 & 84480 & 4455 & 110 & 1 & \\
157329097 & 12 & 12 & 67584 & 4330260 & 32440320 & 61875000 & 43110144 & 13311144 &  2027520 & 160380 & 6600 & 132 & 1 \\
\end{array}
\end{array}
$
}
}
\end{center}
\caption{\parbox[t]{110mm}{%
Left: Number of transitive partial functions on an $11$-set with $k$ fixed points and $l$ out-domain elements
\\
Right: Number of idempotent functions on an $n$-set with $k$ fixed points%
\label{fig:transitive-idempotent-functions}}
}
\end{figure}

Next, we count transitive partial functions, idempotent partial 
functions and idempotent functions on a finite set, where we do
not regard fixed points and out-domain elements.

\begin{lemma}
\label{lm:nb-transitive-idempotent}
Assume $n \in \NAT_{\geq 1}$ and a finite set $X$ with $n = |X|$.
Then we have:
\begin{enumerate}[(a)]
\item
\label{it:transitive-partial-functions}
The number of transitive partial functions on $X$ is
$\sum\limits_{i=0}^n \binom{n}{i} i^{n-i} 2^i$.

\item
\label{it:idemp-partial-functions}
The number of idempotent partial functions on $X$ is
$\sum\limits_{k=0}^n \binom{n}{k} (1+k)^{n-k}$.

\item
\label{it:idempotent-functions}
The number of idempotent functions on $X$ is
$\sum\limits_{k=0}^n \binom{n}{k} k^{n-k}$.
This is also the number of transitive partial functions on $X$ 
with no fixed point.
\end{enumerate}
\end{lemma}

\begin{proof}
(a)
By Theorem \ref{th:transitive-functions} the number of 
transitive partial functions is
$\sum\limits_{k=0}^n \sum\limits_{l=0}^{n-k} \binom{n}{k} \binom{n - k}{l}(k+l)^{n-(k+l)}.$
In the following, we simplify this expression to 
$\sum\limits_{i=0}^n \binom{n}{i} i^{n-i} 2^i$.
\begin{align*}
&~ \sum\limits_{k=0}^n \sum_{l=0}^{n-k} \binom{n}{k} \binom{n - k}{l} (k+l)^{n-(k+l)}
   && \\
= &~ \sum_{k=0}^n \sum_{l=0}^{n-k}  \frac{n!}{k!\, l!\, (n-(k+l))!} (k+l)^{n-(k+l)}
    && \mbox{definition of binomial coefficient} \\
= &~ \sum_{k=0}^n \sum_{l=0}^{n-k} \binom{n}{k+l} \binom{k+l}{k} (k+l)^{n-(k+l)}
    && \mbox{definition of binomial coefficient} \\
= &~ \sum_{k=0}^n \sum_{i=k}^n \binom{n}{i} \binom{i}{k} i^{n-i}
    && \mbox{index shift with $i = k+l$ (see (*) below)} \\
= &~ \sum_{i=0}^n \sum_{k=0}^i \binom{n}{i} \binom{i}{k} i^{n-i}
    && \mbox{permuting sums (see (**) below)} \\
= &~ \sum_{i=0}^n \binom{n}{i} i^{n-i} \sum_{k=0}^i \binom{i}{k}
    && \mbox{distributivity} \\
= &~ \sum_{i=0}^n \binom{n}{i}  i^{n-i} 2^i
    && \mbox{property binomial coefficients} 
\end{align*}
Hint $(*)$:
When shifting the index by defining $i := k+l$,
then we get
$l = 0$ iff $i = k$ and also $l = n-k$ iff $k+l = n$ iff $i = n$.
Hint $(**)$:
It holds
$0 \leq k \leq n$ and $k \leq i \leq n$
iff $0 \leq k \leq i \leq n$ which, in turn, is equivalent
to $0 \leq i \leq n$ and $0 \leq k \leq i$ to hold.

(b)
The number of idempotent partial functions is
$
\sum\limits_{k=0}^n \sum\limits_{l=0}^{n-k}
\binom{n}{k} \binom{n - k}{l} k^{n-(k+l)}
$,
by Corollary~\ref{cor:transitive-functions}\eqref{spec-idempotent}.
Below, we simplify this expression to 
$\sum\limits_{k=0}^n \binom{n}{k} (1+k)^{n-k}$.
By distributivity with properties of subtraction
and the binomial theorem with $(x,y,n) := (1,k,n-k)$ we get:
\[
\sum_{k=0}^n \sum_{l=0}^{n-k} \binom{n}{k} \binom{n - k}{l} k^{n-(k+l)}
= \sum_{k=0}^n \binom{n}{k} \sum_{l=0}^{n-k} \binom{n - k}{l} k^{(n-k)-l}
= \sum_{k=0}^n \binom{n}{k} (1+k)^{n-k}.
\]

(c)
This is a direct consequence of
part (\ref{spec-total}) and part (\ref{spec-no-fp}) of
Corollary~\ref{cor:transitive-functions}.
\end{proof}

\noindent
The bound variable $i$ in the expression 
$\sum\limits_{i=0}^n \binom{n}{i} i^{n-i} 2^i$ of Lemma
\ref{lm:nb-transitive-idempotent}\eqref{it:transitive-partial-functions}
stands for $k+l$ (see the substitution midway in the proof), 
i.e., for the sum of the number of fixed points and the number 
of out-domain elements. 
It can be obtained by choosing $k+l$ elements of $X$, mapping 
the remaining elements to one of the $k+l$ and noting that 
there are $2^{k+l}$ ways to partition the $k+l$ elements in 
the two subsets of fixed points and out-domain elements.
The expression $\sum\limits_{k=0}^n \binom{n}{k} (1+k)^{n-k}$ in Lemma
\ref{lm:nb-transitive-idempotent}\eqref{it:idemp-partial-functions} 
is easily understood by considering the (total) function
$f^t : X \to X \cup\{\bot\}$ that extends $f$:
choose $k$ fixed points from the $n$-set $X$. 
There are then $k+1$ possible values $f^t(x)$ for the 
$n-k$ remaining $x$, namely one of the $k$ fixed points or $\bot$.

The first 23 members of the sequence 
$(\sum\limits_{k=0}^n \binom{n}{k} k^{n-k})_{n \geq 0}$ 
of Lemma~\ref{lm:nb-transitive-idempotent}\eqref{it:idempotent-functions}
(including $n=0$) constitute the sequence A000248 in the OEIS 
data base; see \cite{A000248}.
Furthermore, the table in Figure \ref{fig:nb-trans-idemp-ratios}
displays the number of functions, transitive partial functions, 
idempotent partial functions and idempotent functions on 
an $n$-set, for some $n$ ranging between $1$ and $50$.
The column labeled $\frac{C_2}{C_1}$ shows the ratio of 
the number of transitive partial functions to the number of functions.
This ratio clearly converges to $0$ for increasing $n$; this 
is proved in
Theorem~\ref{ratio-Ptrans-fct}.
The column labeled $\frac{C_3}{C_2}$ shows the ratio of the number 
of idempotent partial functions to the number of transitive partial functions.
It is also clear that it converges to $0$ for increasing $n$; this is proved in Theorem~\ref{th:limits-trans-idemp}\eqref{it:ratio-Pidemp-Ptrans}.
The column labeled $\frac{C_4}{C_3}$ shows the ratio of the number 
of idempotent functions to the number of idempotent partial functions.
This one too seems to converge to $0$; this is proved in Theorem~\ref{th:limits-trans-idemp}\eqref{it:ratio-idemp-Pidemp}.

\begin{figure}
\begin{center}
\scriptsize
$
\begin{array}{r|r|r|r|r|l|l|l}
n &
\multicolumn{1}{c|}{\# \mbox{ functions}} &
\multicolumn{1}{c|}{\# \mbox{ trans. part. fun.}} &
\multicolumn{1}{c|}{\# \mbox{ idemp. part. fun.}} &
\multicolumn{1}{c|}{\# \mbox{ idemp. fun.}} &
\multicolumn{1}{c}{\frac{C_2}{C_1}} &
\multicolumn{1}{c}{\frac{C_3}{C_2}} &
\multicolumn{1}{c}{\frac{C_4}{C_3}}
\\\hline
1 &         1 &       2 &       2 &      1 & 2             & 1            & 0.5          \\
2 &         4 &       8 &       6 &      3 & 2             & 0.75         & 0.5          \\
3 &        27 &      38 &      23 &     10 & 1.40740   & 0.60526 & 0.43478 \\
4 &       256 &     216 &     104 &     41 & 0.84375       & 0.48148 & 0.39423 \\
5 &      3125 &    1402 &     537 &    196 & 0.44864       & 0.38302 & 0.36499  \\
6 &     46656 &   10156 &    3100 &   1057 & 0.21767  & 0.30523 & 0.34096 \\
7 &    823543 &   80838 &   19693 &   6322 & 0.09815 & 0.24361 & 0.32102 \\
8 &  16777216 &  698704 &  136064 &  41393 & 0.04164 & 0.19473 & 0.30421 \\
9 & 387420489 & 6498674 & 1013345 & 293608 & 0.01677 & 0.15593 & 0.28974  \\
10 & 10^{10} &  64579284 & 8076644 & 2237921 & 6.45792 \cdot 10^{-3}  & 0.12506 & 0.27708 \\
20 & 1.04857 \cdot 10^{26} & 9.05586 \cdot 10^{18} & 1.35097 \cdot 10^{17} & 2.72726 \cdot 10^{16} & 8.63634 \cdot 10^{-8}  & 0.01491 & 0.20187 \\
30 & 2.05891 \cdot 10^{44} & 3.70440 \cdot 10^{31} & 7.35951 \cdot 10^{28} & 1.21365 \cdot 10^{28} & 1.79920 \cdot 10^{-13} & 0.00198 & 0.16490 \\
40 & 1.20892 \cdot 10^{64} & 1.48619 \cdot 10^{45} & 4.24955 \cdot 10^{41} & 6.02923 \cdot 10^{40} & 1.22935 \cdot 10^{-19} & 0.00028 & 0.14187 \\
50 & 8.88178 \cdot 10^{84} & 3.39740 \cdot 10^{59} & 1.48297 \cdot 10^{55} & 1.86537 \cdot 10^{54} & 3.82513 \cdot 10^{-26} & 0.00004 & 0.12578 
\end{array}
$
\end{center}
\caption{Number of (transitive, idempotent, partial) functions%
\label{fig:nb-trans-idemp-ratios}}
\end{figure}

We start the investigation of the ratios respectively probabilities
of randomly selected relations and their limiting values 
for the classes we presently discuss with the following result.

\begin{theorem}
\label{ratio-Ptrans-fct}
Assume $n \in \NAT_{\geq 1}$ and a finite set $X$ 
with $n = |X|$.
Then the ratio of the number of transitive partial functions on $X$ to 
the number of functions on $X$ tends 
to $0$ as $n$ tends to infinity.
Using 
Lemma~\ref{lm:nb-transitive-idempotent}(\ref{it:transitive-partial-functions})
and a renaming of the bound variable $i$ to $k$ in it, this is:
\[
\lim\limits_{n \to \infty} 
\frac{\sum\limits_{k=0}^n \binom{n}{k} k^{n-k} 2^k} 
     {n^n} 
= 0
\]
\end{theorem}
\begin{proof}
We assume $n \geq 8$ and calculate
as follows:
\begin{align*}
\frac{\sum\limits_{k=0}^n \binom{n}{k} k^{n-k} 2^k}
     {n^n}
&=    \frac{1}{n^n} \sum_{k=0}^n \binom{n}{k} k^{n-k} 2^k \\
&\leq \frac{1}{n^n} 
      \left(\sum_{k=0}^n \binom{n}{k} \frac{n^{n-1}}{2^n}\right)
        & \mbox{by Lemma \ref{LEMMW-1}} \\
&=    \frac{n^{n-1}}{n^n 2^n} \sum_{k=0}^n \binom{n}{k} 
        & \mbox{distributivity} \\
&=    \frac{n^{n-1}}{n^n 2^n} 2^n 
        & \mbox{property binomial coefficients} \\
&=    \frac{n^{n-1}}{n^n}  \\
&=    \frac{1}{n} 
\end{align*}
Notice, that $n \geq 8$ is not a restriction and 
the shown estimate implies the claimed limiting value, 
since $n$ tends to infinity.
\end{proof}

\noindent
The ratio of this theorem cannot be interpreted as a probability,
since it is not possible to select randomly a partial function on 
$X$ from the set of functions on $X$.
If we take the set of partial function on $X$ instead of the set 
of function on $X$, then the corresponding ratio  becomes
a probability if we take the first set as sample space and
assume again that elements are chosen uniformly at random.

\begin{theorem}
\label{th:limits-transpf-over-pf}
Assume $n \in \NAT_{\geq 1}$ and a finite set $X$ 
with $n = |X|$.
Then the probability that a randomly selected partial 
function on $X$ is transitive is
$$
\frac{\sum\limits_{k=0}^n \binom{n}{k} k^{n-k} 2^k}
     {(n+1)^n}
$$
and this probability converges to 0 if $n$ tends to infinity.
\end{theorem}
\begin{proof}
The first claim follows from
Lemma~\ref{lm:nb-transitive-idempotent}\eqref{it:transitive-partial-functions}
and the fact that $(n+1)^n$ is the number of partial functions on $X$.
The limiting value is an immediate consequence of
$\frac{1}{(n+1)^n} \leq \frac{1}{n^n}$ and Theorem \ref{ratio-Ptrans-fct}.
\end{proof}

\noindent
We now consider three further ratios and show that all of 
them converge to 0 if the cardinality of the carrier
set $X$ tends to infinity.
Also the ratios of the following Theorem \ref{th:limits-trans-idemp} 
can be interpreted as probabilities.
For example, part (a) says that the probability that
a randomly selected transitive partial function on $X$ is
idempotent tends to 0 if $|X|$ tends to infinity.

\begin{theorem}
\label{th:limits-trans-idemp}
Assume $n \in \NAT_{\geq 1}$ and a finite set $X$ 
with $n = |X|$.
Then we have:
\begin{enumerate}[(a)]
\item
\label{it:ratio-Pidemp-Ptrans}
 The ratio of the number of idempotent partial functions on $X$ to 
 the number of transitive partial functions on $X$ tends 
 to $0$ as $n$ tends to infinity.
 That is, using 
 Lemma~\ref{lm:nb-transitive-idempotent}(\ref{it:transitive-partial-functions},\ref{it:idemp-partial-functions}),
 \[
 \lim\limits_{n \to \infty} 
 \frac{\sum\limits_{k=0}^n \binom{n}{k} (1+k)^{n-k}}
      {\sum\limits_{k=0}^n \binom{n}{k} k^{n-k} 2^k} = 0.
 \]
\item
\label{it:ratio-idemp-Pidemp}
 The ratio of the number of idempotent functions on $X$ to the 
 number of idempotent partial functions on $X$ tends to $0$ as $n$ 
 tends to infinity.
 That is, using 
 Lemma~\ref{lm:nb-transitive-idempotent}(\ref{it:idemp-partial-functions},\ref{it:idempotent-functions}),
 \[
 \lim\limits_{n \to \infty} 
 \frac{\sum\limits_{k=0}^n \binom{n}{k} k^{n-k}}
      {\sum\limits_{k=0}^n \binom{n}{k} (1+k)^{n-k}} = 0.
 \]
\item
\label{it:ratio-idemp-Ptrans}
 The ratio of the number of idempotent functions on $X$ to 
 the number of transitive partial functions on $X$ tends 
 to $0$ as $n$ tends to infinity.
 That is, using 
 Lemma~\ref{lm:nb-transitive-idempotent}(\ref{it:transitive-partial-functions},\ref{it:idempotent-functions}),
 \[
 \lim\limits_{n \to \infty} 
 \frac{\sum\limits_{k=0}^n \binom{n}{k} k^{n-k}}
      {\sum\limits_{k=0}^n \binom{n}{k} k^{n-k} 2^k} = 0.
 \]
\end{enumerate}
\end{theorem}
\begin{proof}
For this proof we define
\begin{align*}
	a_n &= \sum_{k=0}^n \binom{n}{k} (k+1)^{n-k},\\
	b_n &= \sum_{k=0}^n \binom{n}{k} k^{n-k},\\	
	c_n &= \sum_{k=0}^n \binom{n}{k} k^{n-k}2^k.
\end{align*}
\begin{enumerate}[(a)]
\item First we show 
\begin{equation}
	\label{eq:bncn}
	\frac{b_n}{c_n}\leq\frac{1}{n^3}
\end{equation}
for all $n\geq 256$ since this implies the claim as $n$ tends to 
infinity, as is shown below. 
In the following, we prove the equivalent property
$b_nn^3-c_n\leq 0$, i.e., 
\begin{equation}
	\label{eq:limit-idemp-over-trans}
	\sum_{k=0}^n \binom{n}{k} k^{n-k} (n^3 - 2^k) \leq 0
\end{equation}
for $n\geq 256$. 
To this end, suppose that $n\geq 256$ and let be the function 
$h_n(k):=k^{n-k}$ from Lemma \ref{LEM2}, and, furthermore, 
define $m := \lfloor \log_2(n^3)\rfloor=\lfloor 3\log_2(n)\rfloor$. 
Then
\begin{equation}
	\label{split}
	\mbox{$n^3- 2^k \geq 0$ if $0 \leq k \leq m$ and $n^3- 2^k < 0$ if $m < k$}.
\end{equation}
We are now ready to prove~\eqref{eq:limit-idemp-over-trans}.
\begin{align*}
& \sum_{k=0}^n \binom{n}{k} k^{n-k} (n^3- 2^k)
\\
=\ & \sum_{k=0}^m \binom{n}{k} k^{n-k} (n^3- 2^k) 
\\
   &  + \sum_{k=m+1}^{2m+1} \binom{n}{k} k^{n-k} (n^3- 2^k) 
\hspace*{28mm}
\parbox[t]{70mm}{This sum is defined, since $\binom{n}{k}$ is
    defined because $2m+1 \leq n$ for $n \geq 30$}
\\
   &   + \sum_{k=2m+2}^n \binom{n}{k} k^{n-k} (n^3- 2^k)
\hspace*{26mm}
\parbox[t]{70mm}{This sum is at most $0$, since $\binom{n}{k} k^{n-k} > 0$ and $n^3- 2^k \leq 0$ for $2m+2 \leq k \leq n$ by ~\eqref{split}}
\\
\leq\ & \sum_{k=0}^m \binom{n}{k} k^{n-k} (n^3- 2^k) 
   + \sum_{k=m+1}^{2m+1} \binom{n}{k} k^{n-k} (n^3- 2^k) 
\\
=\ & \sum_{k=0}^m \binom{n}{k}h_n(k)(n^3- 2^k) 
   + \sum_{k=m+1}^{2m+1} \binom{n}{k}h_n(k)(n^3- 2^k) 
\\
\leq\ & 0,
\end{align*}
where the last step follows from the following consideration. 
First note that there is the same number of expressions in the two sums.
We compare the terms for $i$ from the first and for $m+1+i$ from the 
second sum. For $i=0$, i.e., $k=0$ in the first sum and $k=m+1$ in the second,
we get 
\begin{align*}
& \binom{n}{0} h_n(0) (n^3- 2^0) + \binom{n}{m+1} h_n(m+1) (n^3- 2^{m+1}) \\
& \quad\qquad = 0 + \binom{n}{m+1} h_n(m+1) (n^3- 2^{m+1}) \\
& \quad\qquad \leq 0
\end{align*}
by \eqref{split}. 
Now, let $i \in \{1,\ldots,m\}$. 
Note that $2m+1\leq\lfloor\frac{n}{2}\rfloor$ for $n\geq 74$, and, 
hence, $\binom{n}{i}\leq\binom{n}{m+1+i}$ for $n\geq 74$
since $\binom{n}{i}$ is increasing in the interval $[0,\lfloor\frac{n}{2}\rfloor]$. 
We obtain
\begin{align*}
	\binom{n}{i} h_n(i) (n^3- 2^i)
	&\leq \binom{n}{m+1+i} h_n(i) (n^3- 2^i) && \mbox{$n^3-2^i\geq 0$ and $n\geq 74$}\\	
	&\leq \binom{n}{m+1+i} h_n(m+1+i) (n^3- 2^i) &&\parbox[t]{49mm}{$n^3-2^i\geq 0$ and Lemma \ref{LEM2} since $m+1+i \leq 2m+1 \leq k_n$, i.e., $m+1+i$ is in the interval where $h_n$ increases ($n \geq 256$)}
\end{align*}
Therefore, it suffices to prove
$(n^3-2^i) + (n^3 - 2^{m+1+i}) \leq 0$. Since $m<m+i$ we obtain $n^3 \leq 2^{m+i}$ by \eqref{split}, which implies
$2n^3\leq 2^i + 2^{m+1+i}$, and, hence, $(n^3-2^i) + (n^3 - 2^{m+1+i}) \leq 0$.

After these preparations we are now ready to prove (a), i.e., we have to show $\lim\limits_{n\to\infty}\frac{a_n}{c_n}=0$. For this, we have
\begin{align*}
	a_n
	&= \sum_{k=0}^n \binom{n}{k} (k+1)^{n+1-(k+1)}\\
	&= \sum_{k=0}^n\frac{k+1}{n+1} \binom{n+1}{k+1} (k+1)^{n+1-(k+1)} && \mbox{by Lemma \ref{BINOM}(c)}\\
	&= \sum_{k=1}^{n+1}\frac{k}{n+1} \binom{n+1}{k} k^{n+1-k}\\
	&= \sum_{k=0}^{n+1}\frac{k}{n+1} \binom{n+1}{k} k^{n+1-k}\\
	&= 1+\sum_{k=0}^n\frac{k}{n+1} \binom{n+1}{k} k^{n+1-k} && \frac{n+1}{n+1} \binom{n+1}{n+1} (n+1)^{n+1-(n+1)}=1\\
	&= 1+\sum_{k=0}^n\frac{k}{n+1}\frac{n+1}{n+1-k} \binom{n}{k} k^{n+1-k} && \mbox{by Lemma \ref{BINOM}(b)}\\	
	&= 1+\sum_{k=0}^n\frac{k^2}{n+1-k} \binom{n}{k} k^{n-k}\\
	&\leq 1+\sum_{k=0}^n\frac{n^2}{n+1-n} \binom{n}{k} k^{n-k} && k\leq n\\	
	&= 1+n^2\sum_{k=0}^n \binom{n}{k} k^{n-k}\\
	&= 1+n^2b_n.
\end{align*}
Now suppose $n\geq 256$. Then we have
\begin{align*}
	\frac{a_n}{c_n}
	&\leq \frac{1+n^2b_n}{c_n}\\
	&= \frac{1}{c_n}+n^2\frac{b_n}{c_n}\\
	&\leq \frac{1}{c_n}+\frac{n^2}{n^3} && \mbox{by \eqref{eq:bncn}}\\
	&= \frac{1}{c_n}+\frac{1}{n},
\end{align*}
so that we obtain
\[
\lim_{n\to\infty}\frac{a_n}{c_n}\leq \lim_{n\to\infty}\left(\frac{1}{c_n}+\frac{1}{n}\right)=\lim_{n\to\infty}\frac{1}{c_n}+\lim_{n\to\infty}\frac{1}{n}=0.
\]
\item First we have
\begin{align*}
	b_n
	&= \sum_{k=0}^n \binom{n}{k} k^{n-k}\\
	&= \sum_{k=1}^n \binom{n}{k} k^{n-k} && \binom{n}{0} 0^{n-0}=0\\
	&= \sum_{k=0}^{n-1} \binom{n}{k+1} (k+1)^{n-(k+1)}\\	
	&= \sum_{k=0}^{n-1} \binom{n}{k} \frac{n-k}{k+1}(k+1)^{n-(k+1)} && \mbox{by Lemma \ref{BINOM}(a)}\\	
	&= \sum_{k=0}^{n-1} \binom{n}{k} \frac{n-k}{(k+1)^2}(k+1)^{n-k}\\
	&= \sum_{k=0}^n \binom{n}{k} \frac{n-k}{(k+1)^2}(k+1)^{n-k} && \binom{n}{n} \frac{n-n}{(n+1)^2}(n+1)^{n-n}=0.
\end{align*}
In the following, we want to show that 
\[
\frac{b_n}{a_n}\leq\frac{1}{\ln(n+1)}
\]
for all $n\geq 1453$ since this implies the claim as $n$ tends 
to infinity. 
To this end, we
use the result of the previous derivation and
compute
\begin{align*}
	\frac{b_n}{a_n}\leq\frac{1}{\ln(n+1)}
	&\iff \sum_{k=0}^n \binom{n}{k} (k+1)^{n-k}\left(\ln(n+1)\frac{n-k}{(k+1)^2}-1\right)\leq 0\\
	&\iff \sum_{k=0}^n\frac{1}{(k+1)^2} \binom{n}{k} (k+1)^{(n+1)-(k+1)} \\
	& \phantom{\iff \sum_{k=0}^n}
	    \left(\ln(n+1)(n-k)-(k+1)^2\right)\leq 0\\	
	&\iff \sum_{k=0}^ng_n(k)h_{n+1}(k+1)\left(\ln(n+1)(n-k)-(k+1)^2\right)\leq 0,	
\end{align*}
where $g_n(k):=\frac{1}{(k+1)^2} \binom{n}{k}$ and $h_n$ is the function from Lemma \ref{LEM2}. Therefore, we need to show the equivalent property
\begin{equation}
\label{eq:limit-idemp-func-over-partfunc}
\sum_{k=0}^ng_n(k)h_{n+1}(k+1)\left(\ln(n+1)(n-k)-(k+1)^2\right)\leq 0
\end{equation}
for all $n\geq 1453$. Before we do this we first show that if $n\geq 4$, then $g_n(k)$ is monotone in the interval $[0,\lfloor\frac{n}{2}\rfloor-1]$. To this end, it is sufficient to show that
$g_n(k)\leq g_n(k+1)$ for $0\leq k\leq\lfloor\frac{n}{2}\rfloor-2$. Let $n\geq 4$ and $0\leq k\leq\lfloor\frac{n}{2}\rfloor-2$ and compute
\begin{align*}
	g_n(k)\leq g_n(k+1)
	&\iff \frac{1}{(k+1)^2} \binom{n}{k} \leq\frac{1}{(k+2)^2}\frac{n-k}{k+1} \binom{n}{k} && \mbox{by Lemma \ref{BINOM}(a)}\\
	&\iff (k+2)^2\leq (k+1)(n-k)\\
	&\iff 2k^2+(5-n)k+(4-n)\leq 0.
\end{align*}
If $n\geq 4$, then $n^2-2n-7\geq 0$ so that the last property from above is equivalent to $k\leq \frac{1}{4}(\sqrt{n^2-2n-7}+n-5)$ since $k\geq 0$.
Therefore, it is sufficient to show that we have $\frac{n}{2}-2\leq\frac{1}{4}(\sqrt{n^2-2n-7}+n-5)$. We compute
\begin{align*}
	\frac{n}{2}-2\leq\frac{1}{4}(\sqrt{n^2-2n-7}+n-5)
	&\iff n-3\leq\sqrt{n^2-2n-7}\\
	&\iff (n-3)^2\leq n^2-2n-7\\	
	&\iff -4n+16\leq 0\\
	&\iff n\geq 4.
\end{align*}
Now, we define $m:=\lfloor\frac{1}{2}(\sqrt{\ln(n+1)(4(n+1)+\ln(n+1))}-\ln(n+1)-2)\rfloor$, and then we have
\begin{equation}
\label{split2}
\ln(n+1)(n-k)-(k+1)^2\geq 0\mbox{ if }0\leq k\leq m\mbox{ and }\ln(n+1)(n-k)-(k+1)^2< 0\mbox{ if }m<k
\end{equation}
because $m$ is the positive root of $(k+1)^2-\ln(n+1)(n-k)=k^2+(2+\ln(n+1))k+1-\ln(n+1)n$. 
Before we can prove \eqref{eq:limit-idemp-func-over-partfunc} we now show that $g_n$ and $h_{n+1}$ are increasing up to the bounds $2m+2$ respectively $2m+3$, i.e., that we have
\begin{eqnarray}
\label{eq:smaller-n-div-2}
2m+2\leq\lfloor\frac{n}{2}\rfloor-2\\
\label{eq:smaller-k-n-plus-1}
2m+3\leq\lfloor k_{n+1}\rfloor
\end{eqnarray}
where $k_{n+1}$ is the maximum of the function $h_{n+1}$ (see Lemma \ref{LEM2}) for all $n\geq 1453$. For \eqref{eq:smaller-n-div-2} we obtain
\begin{align*}
	\lefteqn{2m+2\leq\lfloor\frac{n}{2}\rfloor-2}\\
	&\iff 2m+4\leq\frac{n}{2}\hspace*{3em}\mbox{since $2m+4\in\NAT$}\\
	&\iff m\leq\frac{n-8}{4}\\
	&~\Longleftarrow\frac{1}{2}(\sqrt{\ln(n+1)(4(n+1)+\ln(n+1))}-\ln(n+1)-2)\leq\frac{n-8}{4}\\
	&\iff \sqrt{\ln(n+1)(4(n+1)+\ln(n+1))}\leq\frac{n-8}{2}+\ln(n+1)+2\\
	&\iff 2\sqrt{\ln(n+1)(4(n+1)+\ln(n+1))}\leq n+2\ln(n+1)-4\\	
	&\iff 4\ln(n+1)(4(n+1)+\ln(n+1))\leq (n+2(\ln(n+1)-2))^2\\	
	&\iff 16\ln(n+1)n+16\ln(n+1)+4\ln(n+1)^2\\
	&\hspace*{5em}\leq n^2+4(\ln(n+1)-2)n+4(\ln(n+1)-2)^2\\	
	&\iff 0\leq n^2-4(3\ln(n+1)+2)n+4((\ln(n+1)-2)^2-4\ln(n+1)-\ln(n+1)^2)\\
      &\iff 0\leq n^2-4(3\ln(n+1)+2)n-16(2\ln(n+1)-1).
\end{align*}
The last inequality holds for $n\geq 60$. 
Equality holds for approximately $0.4154$ and $59.0990$. 
The expression on the right side is negative between $0.4154$ and $59.0990$
with a minimum approximately at $31.1299$. 
For \eqref{eq:smaller-k-n-plus-1} we have ${\rm e}^{\rm e}\approx 15.1543$ 
so that ${\rm e}^{\rm e}\leq n+1$ for all $n\geq 15$ follows. 
For these $n$ we get 
${\rm e}\leq \ln(n+1)$, and, hence, 
$(n+1){\rm e}\leq (n+1)\ln(n+1)=\ln(n+1){\rm e}^{\ln(n+1)}$. We obtain
$W((n+1){\rm e})\leq\ln(n+1)$ by the definition and monotonicity
of $W$. 
{From} this we conclude 
$\frac{n+1}{\ln(n+1)}\leq\frac{n+1}{W((n+1){\rm e})}=k_{n+1}$ and compute
\begin{align*}
	\lefteqn{2m+3\leq\lfloor k_{n+1}\rfloor}\\
	&\iff 2m+3\leq k_{n+1}\hspace*{3em} \mbox{since $2m+3\in\NAT$}\\	
	&~\Longleftarrow 2m+3\leq \frac{n+1}{\ln(n+1)}\\
	&\iff m\leq\frac{n+1-3\ln(n+1)}{2\ln(n+1)}\\
	&~\Longleftarrow \frac{1}{2}(\sqrt{\ln(n+1)(4(n+1)+\ln(n+1))}-\ln(n+1)-2)\leq\frac{n+1-3\ln(n+1)}{2\ln(n+1)}\\	
	&\iff \ln(n+1)\sqrt{\ln(n+1)(4(n+1)+\ln(n+1))}\leq n+1+\ln(n+1)(\ln(n+1)-1)\\
	&\iff \ln(n+1)^3(4(n+1)+\ln(n+1))\leq (n+1+\ln(n+1)(\ln(n+1)-1))^2\\		
	&\iff 4\ln(n+1)^3n+\ln(n+1)^3(4+\ln(n+1))\\
	&~~~~~~~~~~~~~\leq n^2+2(1+\ln(n+1)(\ln(n+1)-1))n+(1+\ln(n+1)(\ln(n+1)-1))^2\\	
	&\iff 0\leq n^2+2(1+\ln(n+1)(\ln(n+1)-1)-2\ln(n+1)^3)n\\
	&~~~~~~~~~~~~~+(1+\ln(n+1)(\ln(n+1)-1))^2-\ln(n+1)^3(4+\ln(n+1)).
\end{align*}
The last inequality holds for $n \geq 1453$. 
Equality holds for $n$ approximately $1.0478$ and $1452.3706$. 
The expression on the right side is positive for $n\leq 1.0478$ 
with a local minimum at $0$ and a local maximum at approximately $0.4496$ 
and negative in the range from $1.0478\leq n\leq 1452.3706$ with 
a minimum approximately at $825.0153$.
We are now ready to prove~\eqref{eq:limit-idemp-func-over-partfunc}. 
To this end we use the abbreviation 
$t_{n,k}:=g_n(k)h_{n+1}(k+1)\left(\ln(n+1)(n-k)-(k+1)^2\right)$ and compute
\begin{align*}
	\lefteqn{\sum_{k=0}^n t_{n,k}}\\
      &= \sum_{k=0}^m t_{n,k}  + t_{n,m+1} + \sum_{k=m+2}^{2m+2} t_{n,k} + \sum_{k=2m+2}^nt_{n,k}~~~~~\mbox{$2m+2\leq\lfloor\frac{n}{2}\rfloor-2\leq n$ by \eqref{eq:smaller-n-div-2}}\\
      &\leq \sum_{k=0}^m t_{n,k}+\sum_{k=m+2}^{2m+2} t_{n,k}~~~~~\mbox{by \eqref{split2} for $k=m+1$ and $2m+2\leq k\leq n$}\\
      &=  \sum_{k=0}^mt_{n,m-k}+\sum_{k=0}^m t_{n,m+2+k}\\
      &=  \sum_{k=0}^mt_{n,m-k}+t_{n,m+2+k}\\
      &=  \sum_{k=0}^mg_n(m-k)h_{n+1}(m-k+1)\left(\ln(n+1)(n-(m-k))-(m-k+1)^2\right)+t_{n,m+2+k}\\
      &\leq \sum_{k=0}^mg_n(m+2+k)h_{n+1}(m+2+k+1)\left(\ln(n+1)(n-(m-k))-(m-k+1)^2\right)\\
	&\hspace*{4em}+t_{n,m+2+k}\hspace*{2em}\mbox{since $g_n$ and $h_{n+1}$ are monotone using \eqref{eq:smaller-n-div-2} and \eqref{eq:smaller-k-n-plus-1}}\\
	&= \sum_{k=0}^mg_n(m+2+k)h_{n+1}(m+3+k)\\
	&\hspace*{4em}\left(\ln(n+1)(n-(m-k))-(m-k+1)^2\right.\\
	&\hspace*{5em}\left.+\ln(n+1)(n-(m+2+k))-(m+2+k+1)^2\right)
\end{align*}
Furthermore, we have
\begin{align*}
	\lefteqn{\ln(n+1)(n-(m-k))-(m-k+1)^2}\\
	\lefteqn{\hspace*{2em}+\ln(n+1)(n-(m+2+k))-(m+2+k+1)^2}\\
	&= \ln(n+1)((n-(m-k))+(n-(m+2+k)))-(m-k+1)^2-(m+2+k+1)^2\\	
	&= \ln(n+1)(2n-2m-2)-(((m+2)-(k+1))^2+((m+2)+(k+1))^2)\\	
	&= \ln(n+1)(2n-2m-2)-(2(m+2)^2+2(k+1)^2)\\	
	&= 2(\ln(n+1)(n-(m+1))-(m+2)^2-(k+1)^2)\\
	&\leq 2(\ln(n+1)(n-(m+1))-((m+1)+1)^2)\\
	&\leq 0
\end{align*}
where the last $\leq$ follows from \eqref{split2} because $m<m+1$. Finally, we obtain
\[
	\lim_{n\to\infty}\frac{b_n}{a_n}
	\leq \lim_{n\to\infty}\frac{1}{\ln(n+1)}=0.
\]
\item This follows immediately from (a) and (b), or, alternatively, from  \eqref{eq:bncn}.
\end{enumerate}
\end{proof}

\noindent
We now look at how the numbers of transitive partial functions,
idempotent partial functions and idempotent functions change
when the specified number $k$ of fixed points changes. 
For a given $n \in \NAT_{\geq 1}$, an $n$-set $X$ and all 
$k \in \{0,\ldots,n\}$ we define 
\begin{enumerate}[(1)]
\item
\label{it:FPtu}
$\FP^\textit{tu}_n(k) := 
       \binom{n}{k}\sum\limits_{l=0}^{n-k} \binom{n - k}{l}(k+l)^{n-(k+l)}$
      as the number of $t$ransitive partial functions 
      ($u$nivalent relations) on $X$
      that have $k$ fixed points,
\item
\label{it:FPiu}
$\FP^\textit{iu}_n(k) := 
      \binom{n}{k} \sum\limits_{l=0}^{n-k} \binom{n - k}{l} k^{n-(k+l)}$
      as the number of $i$dempotent partial functions 
      ($u$nivalent relations) on $X$
      that have $k$ fixed points, and 
\item
\label{it:FPi}
$\FP^i_n(k) := \binom{n}{k}k^{n-k}$
      as the number of $i$dempotent functions on $X$ that 
      have $k$ fixed points
\end{enumerate}
additionally to the hitherto defined numbers $\FP_n(k)$, $\FP^u_n(k)$, 
$\FP^p_n(k)$ and $\INV_n(k)$ for (partial) functions.
Let $k_n^{\max}$ be such that $\FP^*_n(k_n^{\max})$ is maximal,
with $* \in \{\textit{tu},\textit{iu},i\}$.
The top table in Figure \ref{fig:pos-maximal} displays $k_n^{\max}$ for some 
values of $n$ for the three lists 
$\FP^*_n(0),\ldots, \FP^*_n(n)$. 
For instance, $k_{30}^{\max} = 6$ for $\FP^\textit{tu}_n$.
It may be  that $k_{30}^{\max} = k_{31}^{\max}$.

In the previous sections we could determine precisely where 
the various lists are increasing or decreasing; in other words,
we could determine $k_n^{\max}$.
See again Theorems \ref{FUN-COMP} to \ref{PERM-COMP} of
Section \ref{WITHFP} and Theorem \ref{J_n(k)-increase-decrease}
of Section \ref{WITHOUTFP}.
But the problems with the classes of this section are
much more difficult.
So, we use a different approach. 
The lower table in Figure~\ref{fig:pos-maximal} gives the ratios 
over $n$ of the entries of the top table. 
For instance, $\frac{k_{30}^{\max}}{30} = \frac{6}{30} = 0.2$ for $\FP^\textit{tu}_n$.
\begin{figure}
\begin{center}
\scriptsize
$
\begin{array}{r|rrrrrrrrrrrrrrrrrrr}
n & 1 & 2 & 3 & 4 & 5 & 6 & 7 & 8 & 9 & 10 & 20 & 30 & 40 & 50 & 60 & 70 & 80 & 90 & 100 \\\hline
\FP^\textit{tu}_n & 0 & 1 & 1 & 1 & 1 & 2 & 2 & 2 & 2 &  2 &  4 &  6 &  8 &  9 & 11 & 12 & 14 & 15 &  16 \\
\FP^\textit{iu}_n & 0 & 1 & 1 & 2 & 2 & 3 & 3 & 3 & 4 &  4 &  7 & 10 & 13 & 16 & 18 & 21 & 23 & 26 &  28 \\
\FP^i_n & 1 & 1 & 2 & 2 & 3 & 3 & 3 & 4 & 4 &  4 &  8 & 11 & 14 & 16 & 19 & 21 & 24 & 26 &  29 \\
\end{array}
$

\bigskip

$\begin{array}{r|llllllllll}
n & 1 & 2 & 3 & 4 & 5 & 6 & 7 & 8 & 9 \\\hline
\FP^\textit{tu}_n & 0 & 0.5 & 0.3333 & 0.25 & 0.2 & 0.3333 & 0.2857 & 0.25  & 0.2222 \\
\FP^\textit{iu}_n & 0 & 0.5 & 0.3333 & 0.5  & 0.4 & 0.5 & 0.4285 & 0.375 & 0.4444 \\
\FP^i_n & 1 & 0.5 & 0.6666 & 0.5 & 0.6 & 0.5 & 0.4285 & 0.5 & 0.4444 \\\hline\hline
n & 10 & 20 & 30 & 40 & 50 & 60 & 70 & 80 & 90 & 100 \\\hline
\FP^\textit{tu}_n & 0.2 & 0.2  & 0.2 & 0.2 & 0.18 & 0.1833 &  0.1714 & 0.175 & 0.1666 & 0.16 \\
\FP^\textit{iu}_n & 0.4 & 0.35 & 0.3333 & 0.325 & 0.32 & 0.3 & 0.3 &  0.2875 & 0.2888 & 0.28 \\
\FP^i_n & 0.4 & 0.4 & 0.3666 & 0.35 & 0.32 & 0.3166 & 0.3 & 0.3 & 0.2888 & 0.29 \\
\end{array}
$
\end{center}
\caption{\parbox[t]{109mm}{%
Top array: First value of $k$ for which $\FP^\textit{tu}_n(k)$, 
  $\FP^\textit{iu}_n(k)$ and $\FP^i_n(k)$ are maximal \\
Bottom array: ratio of lines $2$, $3$ and $4$ of the top array over line $1$%
\label{fig:pos-maximal}}
}
\end{figure}
All these ratios slowly decrease.
Using the technique of least square fitting to find the line 
that best approximates the $k_n^{\max}$, we get lines 
$an + b$ with the slopes $a$ and initial ordinates $b$ listed 
in the following table, for four intervals.
\[
\begin{array}{r|c|c|c|c|c|c|c|c|}
\multicolumn{1}{c}{}
& \multicolumn{2}{c}{1 \leq n \leq 100}
& \multicolumn{2}{c}{10 \leq n \leq 40} 
& \multicolumn{2}{c}{40 \leq n \leq 70} 
& \multicolumn{2}{c}{70 \leq n \leq 100} \\
\cline{2-9}
& a & b & a & b & a & b & a & b
\\\cline{2-9}
\FP^\textit{tu}_n & 0.15516 & 1.32970 & 0.17642 & 2.54194 & 0.15083 & \phantom{0}7.67957 & 0.13838 & 12.12688 \\
\FP^\textit{iu}_n & 0.27316 & 1.91861 & 0.29744 & 4.18710 & 0.26363 & 13.07742 & 0.25072 & 20.86452 \\
\FP^i_n & 0.27420 & 2.36693 & 0.30723 & 4.51183 & 0.25918 & 13.67527 & 0.25095 & 21.36129 \\
\cline{2-9}
\end{array}
\]
The lower table in Figure~\ref{fig:pos-maximal} shows that the 
ratio $\frac{k_n^{\max}}{n}$ tends to decrease with increasing 
$n$.
Local increases also happen, as the table shows. 
The slopes $a$ in the previous table show the same trend in 
a more synthetic way. 
Both tables also show that $k_n^{\max}$ for the list
$\FP^\textit{iu}_n(0),\ldots,\FP^\textit{iu}_n(n)$ is almost the 
same as that for the list $\FP^\textit{i}_n(0),\ldots,\FP^\textit{i}_n(n)$.
This is not surprising, since $(1+k)^{n-k}$ is quite 
similar to $k^{n-k}$.

Now, the question is whether
$\lim\limits_{n \to \infty} \frac{k_n^{\max}}{n} = 0$ or not for
each of the seven classes of (partial) functions we have 
considered so far.
This holds for the list $\FP_n(0) ,\ldots, \FP_n(n)$, 
since $k_n^{\max} = 1$ by Theorem~\ref{FUN-COMP}.
It also holds for the list
$\FP^u_n(0) ,\ldots, \FP^u_n(n)$, 
since in this case $k_n^{\max} = 0$ by Theorem~\ref{PFUN-COMP},
for the list $\FP^p_n(0) ,\ldots, \FP^p_n(n)$, 
since here $k_n^{\max} = 0$ or $k_n^{\max} = 1$ by
Theorem~\ref{PERM-COMP}, depending on the parity of $n$,
and for the list $\INV_n(0) ,\ldots, \INV_n(n)$, 
since $k_n^{\max} \approx \sqrt{n}$ by Theorem~\ref{J_n(k)-increase-decrease}. 
So far, we have no proof that
the limiting value is $0$ also 
for the lists given by
\eqref{it:FPtu}, \eqref{it:FPiu} and \eqref{it:FPi}
above. 
Because $\frac{\sqrt{n}}{n}$ decreases faster than $\frac{k_n^{\max}}{n}$ 
for any of the
three lists, it cannot be used to provide an upper bound which would 
lead to the conclusion that 
$\lim\limits_{n \to \infty} \frac{k_n^{\max}}{n} = 0$.


\section{A Result on Partial Functions With At Most One Fixed Point}
\label{Sec:PFunBij}

Throughout this section we assume a fixed non-empty set $X$.
We define $\textsl{Fun\/}(X)$ to be the set of all 
(total) functions on $X$,
$\textsl{Pfun\/}(X)$ to be the set of all partial functions on $X$,
$\textsl{Pfun\/}_0(X)$ to be the set of all partial functions on $X$ without a 
fixed point and
$\textsl{Pfun\/}_1(X)$ to be the set of all partial functions on $X$ with precisely 
one fixed point.
It is well known that if $X$ is finite and $|X| = n$, then
$|\textsl{Fun\/}(X)|=n^n$. 
In Theorem \ref{PFUN-COMP} we have shown that in this case the sets
$\textsl{Pfun\/}_0(X)$ and $\textsl{Pfun\/}_1(X)$ also have $n^n$ elements, 
i.e., we have 
$$
n^n = |\textsl{Fun\/}(X)| 
    = |\textsl{Pfun\/}_0(X)| 
    = |\textsl{Pfun\/}_1(X)| .
$$
As a consequence, if $X$ is finite then there are bijections between the three sets
$\textsl{Fun\/}(X)$, $\textsl{Pfun\/}_0(X)$ and $\textsl{Pfun\/}_1(X)$.
In the remainder of the section we will prove that this even holds without
the finiteness assumption, i.e., the statement from above can be generalized 
to $X$ being infinite. 

We start with the construction of a bijection between the two sets 
$\textsl{Pfun\/}_0(X)$ and $\textsl{Pfun\/}_1(X)$.
To this end, we first consider transpositions.
For two elements $b,c \in X$ the transposition $\transp{b}{c}$ is a bijective 
function on $X$ defined by 
$$ 
\transp{b}{c}(x) =
\left\{\begin{array}{ll}c & \mbox{if }x=b\\ 
                        b &\mbox{if }x=c\\ 
                        x &\mbox{otherwise,}
       \end{array}
\right .
$$
for all $x \in X$.
Obviously, we have 
$(\transp{b}{c} \circ\transp{b}{c})(x) = \transp{b}{c}(\transp{b}{c}(x)) = x$,
for all $x \in X$.
Besides $X$, in what follows we also fix an arbitrary element $a \in X$. 
Then we define the subsets $M_{\rm undef}$ and $M_b$, for all $b \in X$ 
with $a \ne b$, of $\textsl{Pfun\/}_0(X)$ as follows:
\begin{align*}
 M_{\rm undef} &:= \{f\mid f\in\textsl{Pfun\/}_0(X)\wedge a\not\in{\rm dom}(f)\}\\
 M_b           &:= \{f\mid f\in\textsl{Pfun\/}_0(X)\wedge a\in{\rm dom}(f)\wedge f(a)=b\}
\end{align*}
The collection of sets from above forms a partition of the set $\textsl{Pfun\/}_0(X)$ 
because each partial function on $X$ with no fixed point is either undefined 
for $a$ or $f(a) \ne a$ holds. 
Furthermore, we define the following subsets $N_b$, for all $b \in X$,
of the set $\textsl{Pfun\/}_1(X)$:
$$ 
N_b := \{g \mid g\in\textsl{Pfun\/}_1(X)\wedge b\in{\rm dom}(g)\wedge g(b)=b\}
$$
Similar to the case of $\textsl{Pfun\/}_0(X)$, the collection of sets 
defined above forms a partition of the set $\textsl{Pfun\/}_1(X)$.
The next step in the construction of a bijection between 
$\textsl{Pfun\/}_0(X)$ and $\textsl{Pfun\/}_1(X)$ consists of the
following auxiliary result.

\begin{lemma}
\label{Lem:Nbij}
For all $b, c \in X$ with $b\ne c$ let 
$\chi_{b,c} : \textsl{Pfun\/}(X) \to \textsl{Pfun\/}(X)$ be the function defined by
$$ 
\chi_{b,c}(g) = \transp{b}{c} \circ g \circ \transp{b}{c},
$$
for all $g \in \textsl{Pfun\/}(X)$.
Then the restriction of $\chi_{b,c}$ to the set $N_b$ is a bijective
function from the set $N_b$ to the set $N_c$.
\end{lemma}

\begin{proof}
We first prove $\chi_{b,c}(N_b) \subseteq N_c$ such that the
restriction is in fact a function from $N_b$ to $N_c$.
So, assume $g \in N_b$, i.e., $g$ has $b$ as its unique fixed point. 
Then $\chi_{b,c}(g)$ has $c$ as a fixed point because
\begin{align*}
  \chi_{b,c}(g)(c)
  &= (\transp{b}{c}\circ g \circ\transp{b}{c})(c)\\
  &= (\transp{b}{c}\circ g)(b)\\
  &= \transp{b}{c}(b) & \mbox{$b$ fixed point of $g$}\\
  &= c.
\end{align*}
This is also the only fixed point of $\chi_{b,c}(g)$ because if 
$x\ne c$ would be another fixed point of $\chi_{b,c}(g)$, then
\begin{align*}
  \transp{b}{c}(x)
  &= \transp{b}{c}(\chi_{b,c}(g)(x)) && \mbox{$x$ fixed point of $\chi_{b,c}(g)$}\\
  &= (\transp{b}{c}\circ\transp{b}{c}\circ g \circ\transp{b}{c})(x)\\  
  &= (g \circ\transp{b}{c})(x) && \mbox{$\transp{b}{c}$ transposition}\\
  &= g(\transp{b}{c}(x)),
\end{align*}
i.e., $\transp{b}{c}(x)$ would be a fixed point of $g$. 
But since $x\ne c$ we have $\transp{b}{c}(x)=c$ if $x=b$ or $\transp{b}{c}(x)=x$ 
if $x\ne b$. 
In both cases this would be another fixed point of $g$ different of $b$, a 
contradiction. 
This verifies that $\chi_{b,c}(g)\in N_c$. A similar argument shows that $\chi_{b.c}(f)\in N_b$ for every $f\in N_c$.
Consequently, that the restriction of $\chi_{b,c}$ to $N_b$ is bijective immediately follows from
$$ 
\chi_{b,c}(\chi_{b,c}(g)) 
= \transp{b}{c}\circ\transp{b}{c}\circ g \circ\transp{b}{c}\circ\transp{b}{c} 
= g ,
$$
for all $g \in N_b$, since $\transp{b}{c}$ is a transposition.
\end{proof}

\noindent
As the decisive step in the construction of a bijection between 
the sets $\textsl{Pfun\/}_0(X)$ and $\textsl{Pfun\/}_1(X)$ 
we now define a function $\varphi_a : \textsl{Pfun\/}_0(X)\to N_a$ by
$\textrm{dom}(\varphi_a(f)) = \{a\} \cup \textrm{dom}(f)$ and
$$ 
\varphi_a(f)(x) 
= \left\{\begin{array}{ll}
         a & \mbox{if $x=a$} \\
         f(x) & \mbox{if $x\ne a$ and $x \in {\rm dom}(f)$},
       \end{array}\right. 
$$
for all $f \in \textsl{Pfun\/}_0(X)$ and
$x \in \{a\} \cup \textrm{dom}(f)$.
In order to verify that, given any $f \in \textsl{Pfun\/}_0(X)$,
the partial function $\varphi_a(f)$ is indeed in the set $N_a$, we
use that $\varphi_a(f)$ has $a$ as a fixed point by definition. 
If $x \ne a$ would be another fixed point of $\varphi_a(f)$, then 
we get $x \in {\rm dom}(f)$ and 
$x = \varphi_a(f)(x) = f(x)$, again by the definition of $\varphi_a(f)$.
This is a contradiction to the fact that $f \in \textsl{Pfun\/}_0(X)$, i.e., 
to the fact that $f$ does not have any fixed points. 
Consequently, we have $\varphi_a(f)\in N_a$.

\begin{lemma}
\label{Lem:MNbij}
For the function $\varphi_a : \textsl{Pfun\/}_0(X)\to N_a$ the following properties
hold:
\begin{enumerate}
\item[(a)] Its restriction to the set $M_{\rm undef}$ is a 
           bijective function from $M_{\rm undef}$ to the set $N_a$.
\item[(b)] For all $b \in X$ with $a \ne b$ the restriction of $\varphi_a$ to 
           the set $M_b$ is a bijective function from $M_b$ to the set $N_a$.
\end{enumerate}
\end{lemma}

\begin{proof}
(a) To prove injectivity, we assume that $\varphi_a(f) = \varphi_a(g)$ for 
two functions $f,g$ in $M_{\rm undef}$. 
Then we have
$$
{\rm dom}(f)\cup\{a\} = {\rm dom}(\varphi_a(f)) 
                      = {\rm dom}(\varphi_a(g)) 
                      = {\rm dom}(g)\cup\{a\} 
$$
because of the definition of ${\rm dom}(\varphi_a(f))$ and ${\rm dom}(\varphi_a(g))$.
Since $f,g \in M_{\rm undef}$ we obtain ${\rm dom}(f) = {\rm dom}(g)$. 
If $x \in {\rm dom}(f) = {\rm dom}(g)$ and, therefore, $x \ne a$, then we 
have 
$f(x) = \varphi_a(f)(x) = \varphi_a(g)(x) = g(x)$. 
This implies $f = g$ and, hence, that $\varphi_a$ is injective. 
To prove surjectivity, we suppose an arbitrary partial function $h \in N_a$
to be given.
We define a partial function $f$ on $X$ by ${\rm dom}(f) = {\rm dom}(h) \setminus\{a\}$ 
and $f(x) = h(x)$, for all $x \in {\rm dom}(f)$.
Then we get $f \in M_{\rm undef}$, since $a \not\in {\rm dom}(f)$, and 
that $f$ has no fixed point, since $f(x) = h(x)$, for all $x \in {\rm dom}(f)$, and 
$h$ has $a$ as unique fixed point.
Furthermore, we have 
$$
{\rm dom}(\varphi_a(f)) = {\rm dom}(f) \cup \{a\} 
                        = ({\rm dom}(h)\setminus\{a\}) \cup \{a\}
                        = {\rm dom}(h)
$$ 
by the definition of ${\rm dom}(\varphi_a(f))$ and ${\rm dom}(f)$ and
since $a \in {\rm dom}(h)$. 
If $x \in {\rm dom}(\varphi_a(f)) = {\rm dom}(h)$, we obtain 
$\varphi_a(f)(x) = f(x) = h(x)$, for $x \ne a$, and $\varphi_a(f)(a) = a = h(a)$,
since $a$ is a fixed point of $h$.
This verifies that $\varphi_a(f) = h$.
So, the function $\varphi_a$ is also surjective.

(b) Let a $b \in X$ with $a \ne b$ be given.
To prove injectivity, we assume again that $\varphi_a(f) = \varphi_a(g)$ 
for two functions $f,g \in M_b$.
Then we get
$$
{\rm dom}(f) = {\rm dom}(f) \cup \{a\}
             = {\rm dom}(\varphi_a(f))
             = {\rm dom}(\varphi_a(g))
             = {\rm dom}(g) \cup \{a\}
             = {\rm dom}(g)
$$
by the definition of ${\rm dom}(\varphi_a(f))$ and ${\rm dom}(\varphi_a(g))$ and 
as $a \in {\rm dom}(f)$ and $a \in  {\rm dom}(g)$. 
If $x \in {\rm dom}(f) = {\rm dom}(g)$ and $x \ne a$, then we have 
$f(x) = \varphi_a(f)(x) = \varphi_a(g)(x) = g(x)$. 
Since $f(a) = b = g(a)$, this implies $f = g$.
As a consequence, $\varphi_a$ is injective. 
Now, we prove surjectivity.
So, suppose an arbitrary $h \in N_a$.
We define a partial function $f$ on $X$ by 
${\rm dom}(f) = {\rm dom}(h)$ and $f(x) = h(x)$, for all $x \in {\rm dom}(f)$ with 
$x \ne a$, and $f(a) = b$.
Then we have 
$$
{\rm dom}(\varphi_a(f)) = {\rm dom}(f)\cup\{a\} 
                        = {\rm dom}(f) 
                        = {\rm dom}(h)
$$
by definition of ${\rm dom}(\varphi_a(f))$ and due to
$a \in {\rm dom}(f)$ and ${\rm dom}(f) = {\rm dom}(h)$.
If $x \in {\rm dom}(\varphi_a(f)) = {\rm dom}(h)$, we obtain 
$\varphi_a(f)(x) = f(x) = h(x)$, in case that $x \ne a$, and $\varphi_a(f)(a) = a = h(a)$,
since $a$ is a fixed point of $h$.
Altogether, we have $\varphi_a(f) = h$ and this implies that $\varphi_a$ is also surjective.
\end{proof}

\noindent
After these preparations we are now ready to verify the existence of the
bijection between $\textsl{Pfun\/}_0(X)$ and $\textsl{Pfun\/}_1(X)$.

\begin{theorem}
\label{Th:PFun0=PFun1}
The function $\Psi : \textsl{Pfun\/}_0(X) \to \textsl{Pfun\/}_1(X)$ defined by
$$ 
\Psi(f) = 
\left\{\begin{array}{ll}
       \varphi_a(f) & \mbox{if $f \in M_{\rm undef}$}\\ 
       (\chi_{a,b}\circ\varphi_a)(f) & \mbox{if $f\in M_b$},
      \end{array}\right. 
$$
for all $f \in \textsl{Pfun\/}_0(X)$ and $b \in X$,
is bijective.
\end{theorem}

\begin{proof}
As already mentioned, the collection of the sets $M_{\rm undef}$ and $M_b$, 
where $b \in X \setminus \{a\}$, forms a partition of the set $\textsl{Pfun\/}_0(X)$. 
Since the image of the restriction of the function $\varphi_a$ to the set $M_{\rm undef}$ 
is a subset of the set $N_a$ and $N_a \subseteq \textsl{Pfun\/}_1(X)$ and
for all $b \in X \setminus \{a\}$ the image of the restriction of the composition 
$\chi_{a,b} \circ \varphi_a$ to the set $M_b$ is a subset of the set 
$N_b$ and $N_b \subseteq \textsl{Pfun\/}_1(X)$, we get that the image of $\Psi$ 
is indeed included in the set $\textsl{Pfun\/}_1(X)$. 

By Lemma \ref{Lem:MNbij}(a) the restriction of the function $\varphi_a$ to the set 
$M_{\rm undef}$ is a bijective function from this set to the set $N_a$. 
For all $b \in X \setminus \{a\}$, 
furthermore, by Lemma \ref{Lem:Nbij} and Lemma \ref{Lem:MNbij}(b)
the restriction of the function $\chi_{a,b} \circ \varphi_a$ to the set $M_b$
is a bijection between the sets $M_b$ and $N_b$. 
We also have mentioned that the collection of the sets $N_b$, where $b \in X$, forms
a partition of the set $\textsl{Pfun\/}_1(X)$.
{From} this we get that $\Psi$ is a bijective function from 
$\textsl{Pfun\/}_0(X)$ to $\textsl{Pfun\/}_1(X)$ as claimed.
\end{proof}

\noindent
We want to illustrate the previous theorem in an example where we 
represent relations by Boolean matrices. 
For this example suppose $X = \{1,2,3,4\}$ is a set with $4$ elements.
A relation $R$ on $X$ can be represented by a Boolean matrix 
$(a_{ij})_{i,j\in\{1,2,3,4\}}$, i.e., by a matrix for which the entries 
$a_{ij}$ can take the values $0$ and $1$.
If an entry $a_{ij}=0$, then the pair $(i,j)\not\in R$, and, 
if $a_{ij}=1$, then $(i,j)\in R$. 
Such a matrix is a partial function iff there is at most one non-zero entry in each row. 
An entry on the diagonal of a matrix represents a fixed point. 
Consequently, for a partial function with zero resp.\ one fixed point the 
corresponding matrix has no resp.\ one 
non-zero entry on the diagonal.
Now, we choose $a = 1$ and obtain the $4$ different kinds of matrices 
representing the elements of $M_{\rm undef}$, $M_2$, $M_3$, and $M_4$ 
within $\textsl{Pfun\/}_0(X)$,
i.e., $P_1\in M_{\rm undef}$, $P_2\in M_2$, $P_3\in M_3$, and $P_4\in M_4$.
As mentioned above the values $a_{ij}$ need
to be chosen so that there 
is at most one non-zero entry in each row, of course.
\[
\begin{array}{c@{}c@{}c@{}c}
P_1 & P_2 & P_3 & P_4 \\
\begin{pmatrix}
0 & 0 & 0 & 0 \cr
a_{21} & 0 & a_{23} & a_{24} \cr
a_{31} & a_{32} & 0 & a_{34} \cr
a_{41} & a_{42} & a_{43} & 0 \cr
\end{pmatrix}
&
\begin{pmatrix}
0 & 1 & 0 & 0 \cr
a_{21} & 0 & a_{23} & a_{24} \cr
a_{31} & a_{32} & 0 & a_{34} \cr
a_{41} & a_{42} & a_{43} & 0 \cr
\end{pmatrix}
&
\begin{pmatrix}
0 & 0 & 1 & 0 \cr
a_{21} & 0 & a_{23} & a_{24} \cr
a_{31} & a_{32} & 0 & a_{34} \cr
a_{41} & a_{42} & a_{43} & 0 \cr
\end{pmatrix}
&
\begin{pmatrix}
0 & 0 & 0 & 1 \cr
a_{21} & 0 & a_{23} & a_{24} \cr
a_{31} & a_{32} & 0 & a_{34} \cr
a_{41} & a_{42} & a_{43} & 0 \cr
\end{pmatrix}
\end{array}
\]
Applying the function $\varphi_a$ to each of these partial
functions
modifies the first row of the matrix yielding the following matrix in all $4$ cases:
\[
\begin{array}{c}
Q\\
\begin{pmatrix}
1 & 0 & 0 & 0 \cr
a_{21} & 0 & a_{23} & a_{24} \cr
a_{31} & a_{32} & 0 & a_{34} \cr
a_{41} & a_{42} & a_{43} & 0 \cr
\end{pmatrix}
\end{array}
\]
In order to obtain the final result, we need to apply permutations 
in all cases except for matrices of the form $P_1$. 
In those cases we have to permute
the element $a=1$ with the image of $a$ in $P_i$, i.e., we have to 
swap the first row and column with the $i$-th row and column.
This leads to the following matrices:
\[
\begin{array}{c@{}c@{}c@{}c}
R_1 & R_2 & R_3 & R_4 \\
= Q & \mbox{Switch 1,2 of $Q$} & \mbox{Switch 1,3 of $Q$} & \mbox{Switch 1,4 of $Q$} \\
\begin{pmatrix}
1 & 0 & 0 & 0 \cr
a_{21} & 0 & a_{23} & a_{24} \cr
a_{31} & a_{32} & 0 & a_{34} \cr
a_{41} & a_{42} & a_{43} & 0 \cr
\end{pmatrix}
&
\begin{pmatrix}
0 & a_{21} & a_{23} & a_{24} \cr
0 & 1 & 0 & 0 \cr
a_{32} & a_{31} & 0 & a_{34} \cr
a_{42} & a_{41} & a_{43} & 0 \cr
\end{pmatrix}
&
\begin{pmatrix}
0 & a_{32} & a_{31} & a_{34} \cr
a_{23} & 0 & a_{21} & a_{24} \cr
0 & 0 & 1 & 0 \cr
a_{43} & a_{42} & a_{41} & 0 \cr
\end{pmatrix}
&
\begin{pmatrix}
0 & a_{42} & a_{43} & a_{41} \cr
a_{24} & 0 & a_{23} & a_{21} \cr
a_{34} & a_{32} & 0 & a_{31} \cr
0 & 0 & 0 & 1 \cr
\end{pmatrix}
\end{array}
\]
These four matrices now represent the matrices of the elements from the sets 
$N_i$ for $i\in\{1,2,3,4\}$. Obviously, each of these
matrices has exactly one fixed point since there is exactly 
one non-zero entry on the diagonal. A quick calculation also verifies that these
matrices are partial functions, i.e., only have one non-zero entry in each row, 
if the original matrices $R_1$, $R_2$, $R_3$, and $R_4$ 
also satisfy this property.

By a combination of Theorem \ref{Th:PFun0=PFun1} and the following theorem
we complete the proof that there are bijections between the three sets 
$\textsl{Fun\/}(X)$, $\textsl{Pfun\/}_0(X)$ and $\textsl{Pfun\/}_1(X)$.

\begin{theorem}
\label{Th:Fun=PFun0}
The function $\Phi : \textsl{Fun\/}(X) \to \textsl{Pfun\/}_0(X)$, defined by
$\dom{\Phi(f)} = \{x \in X \mid f(x) \ne x\}$ and 
$\Phi(f)(x) = f(x)$,
for all $f \in \textsl{Fun\/}(X)$ and $x \in\dom{\Phi(f)}$,
is bijective.
\end{theorem}

\begin{proof}
By the definition of $\Phi$ we have for all 
$f \in \textsl{Fun\/}(X)$ and all $x \in X$ the following two properties:
$$
\mbox{(a)~~~}
\Phi(f)(x) = f(x) \iff f(x) \not= x
\quad\qquad
\mbox{(b)~~~}
x \notin \textsl{dom\/}(\Phi(f)) \iff f(x) = x
$$
This immediately implies that the function $\Phi$ is well-defined, since for all
functions $f \in \textsl{Fun\/}(X)$ the partial function $\Phi(f) \in \textsl{Pfun\/}(X)$
has no fixed point.

Now, assume $f,g \in \textsl{Fun\/}(X)$ such that $\Phi(f) = \Phi(g)$.
Furthermore, let $x \in X$.
If $x \in\textsl{dom\/}(\Phi(f)) = \textsl{dom\/}(\Phi(g))$, then
(b) implies $f(x) \neq x$ and $g(x) \neq x$, and then (a) that 
$$
f(x) = \Phi(f)(x) = \Phi(g)(x) = g(x).
$$
If $x \notin \textsl{dom\/}(\Phi(f)) = \textsl{dom\/}(\Phi(g))$, i.e., 
$x \not\in \textsl{dom\/}(\Phi(g))$, then applying (b) twice yields 
$f(x) = x = g(x)$.
This verifies that $f = g$ and, hence, that $\Phi$ is injective.
Now, assume an arbitrary $g \in \textsl{Pfun\/}_0(X)$ to be given.
We define a function $f \in \textsl{Fun\/}(X)$ by
$$
f(x) = \left \{ \begin{array}{cl}
                 g(x) & \mbox{if } x \in \textsl{dom\/}(g) \\
                 x    & \mbox{if } x \notin \textsl{dom\/}(g) ,
                \end{array}
        \right .
$$
for all $x \in X$.
We then have that $x \in\dom{\Phi(f)}$ is equivalent to $f(x) \neq x$ by (b),
which is itself equivalent to $x \in \dom{g}$ by the
definition of $f$ and the fact that $g$ does not have any fixed points. 
Consequently, we have $\dom{\Phi(f)} = \dom{g}$.
If $x \in \dom{\Phi(f)} = \dom{g}$, then we get 
$\Phi(f)(x) = f(x) = g(x)$ using (a) and (b) and the definition of $f$. 
This shows that $\Phi(f) = g$ and, hence, the function $\Phi$ is also surjective.
\end{proof}


\section{Concluding Remarks, Related and Future Work}
\label{CONCL}

For some important classes of relations we have investigated 
the probability of a randomly selected relation from a certain
class to have a reflexive point respectively a fixed point if the
class consists of (partial) functions only.
We also have calculated the limiting values of the probabilities.
The decisive parts for obtaining the probabilities
have been the counting of all relations of the class
under consideration and also of those which have at least one
reflexive point (fixed point for functions).
In this regard we have taken a most general approach.
For each class of relations we not only have counted the relations that have at 
least one reflexive/fixed point but those with a specified number 
$k$ of such points.
If reasonable, we also allowed relations to be heterogeneous.
And, apart from the classes of Section \ref{IDEMP}, we also have shown 
how the numbers of the lists we have obtained
for the classes are related to each other with respect to their sizes.

It seems that there are only a few known results concerning the number 
of relations -- that are not demanded to be functions -- with a 
specified number $k$ of reflexive points.
For example, it is known that on an $n$-set there exist $2^{n(n-1)}$
irreflexive relations and also $2^{n(n-1)}$ reflexive relations.
These results are specific instances of part (d) of
Theorem \ref{THEEXIS-GEN} for $X = Y$, hence $m = n$, and 
$k = 0$ respectively $k = n$.

In contrast to this, there exist many results concerning the number 
of functions with fixed points.
To our knowledge all of them consider functions on a single set only
and mostly functions are counted which have either no or at 
least one fixed point and not a specified number $k$ of fixed points.
We are not aware of results concerning partial functions.

In view of Section \ref{WITHFP} it is known that on an $n$-set 
there are $n^n - (n-1)^n$ functions with at least one fixed point.
This is a specific instance of part (d) of Theorem 
\ref{THEEXIS} for $X = Y$, hence $m = n$.
Also the limiting value $1 - \frac{1}{\textrm{e}}$ of part (a) of
Theorem \ref{THEEXIS-limits} is known and, e.g., in \cite{Precht}
illustrated by means of the following example:
A person writes $n$ letters to $n$ persons and also prepares 
the corresponding $n$ envelopes. 
Then it takes a randomly selected letter, puts it into a randomly 
selected envelope and repeats this until all envelopes contain 
a letter. 
The probability that at least one envelope contains the correct 
letter is approximately $63 \%$.

Given $n \in \NAT_{\geq 1}$ and $k \in \{0,\ldots,n\}$, it is also 
known how many permutations on the set $\{1,\ldots,n\}$ have $k$ 
fixed points.
In the literature such permutations are also called partial 
derangements and their numbers are the rencontres numbers \cite{WikiRecontres}.
The latter equal our numbers $\FP^p_n(k)$.
We have included the proof of Theorem \ref{THEPERM-GEN} to make 
the paper self-contained as much as possible.
It seems a result similar to our Theorem \ref{PERM-COMP} has not 
been published until now.

When strengthening in Section \ref{WITHFP} the class of functions $f : X \to Y$, 
where $X$ is an $m$-set, $Y$ is an $n$-set and $X \subseteq Y$,
we at once changed to the class of permutations on $X$ and left
out the intermediate classes of surjective functions 
$f : X \to Y$ and injective functions $f : X \to Y$.
The first class yields no additional results, since in case 
$X \subset Y$ it is empty and in case $X = Y$ it is equal to 
the class of permutations on $X$.
Also the second class becomes the class of permutations on $X$
if $X = Y$.
For $X \subseteq Y$ it is known that there are
$\frac{n!}{(n-m)!}$ injective functions $f : X \to Y$.
As we submitted our paper we neither knew how many of them have
a specified number $k$ of fixed points nor how many of them have
at least one fixed point and what is the probability that a 
randomly selected injective function $f : X \to Y$ has at least
one fixed point.
In the Fall of 2025 we have been able to show, using the principle of inclusion and exclusion, that there are precisely
$$
\binom{m}{k} \sum_{j=0}^{m-k} (-1)^j 
                              \binom{m-k}{j} \frac{(n-k-j)!}{(n-m)!}
$$
injective function $f : X \to Y$ that have precisely $k$ fixed points,
where $k \in\{0,\ldots,m\}$.
This also allowed us to determine how many injective function $f : X \to Y$ 
have at least one fixed point and the probability that a randomly selected function
has at least one fixed point. Shortly after this we discovered the eprint \cite{Luciano} on the arXiv archive
that shows the same result by even using the same method.

Concerning Section \ref{WITHOUTFP}, it is known that $\INV_n(k)$
is the number of those matchings of the Kuratowski-graph $K_n$
which consist of $k$ edges.
Each such matching $M$ leads to an involution $f$ on 
$\{1,\ldots,n\}$, where $f(x) = y$, if $x$ is contained
in an edge $\{x,y\}$ of $M$, and $f(x) = x$, if $x$ is not
contained in an edge of $M$.
This mapping even establishes a 1-1 correspondence which,
in principle, we have used to prove Theorem \ref{THENUMBPINV}.
Again,
it seems results similar to our 
Theorems \ref{J_n(k)-increase-decrease} and \ref{th:limits-involutions}
have not been published so far.

With regard to Section \ref{IDEMP} it should be mentioned that
the numbers $\binom{n}{k} k^{n-k}$ are known in combinatorics as 
idempotent numbers.
It seems to be unknown how these numbers are related to each 
other with respect to their sizes.
We also are not aware of investigations concerning the number 
of idempotent partial functions and transitive partial functions 
and of such which have a specified number $k$ of fixed points respectively
out-domain elements.

Many results can easily be obtained from those of the previous sections by considering dualities based on the transpose $\Transp{R}$ of a relation $R$.
These dualities are that
\begin{enumerate}
\item
$R$ is a partial function iff $\Transp{R}$ is injective;

\item
$R$ is total iff $\Transp{R}$ is surjective;

\item
$R$ is transitive iff $\Transp{R}$ is transitive;

\item
the fixed points of $R$ are the same as those of $\Transp{R}$.
\end{enumerate}
For an example of such a result, consider $X$ and $Y$ as in
Theorem~\ref{THEEXIS}.
The dual of Theorem~\ref{THEEXIS}(b) is then
that the number of injective relations $R \subseteq Y \times X$ with a fixed point is 
$(n+1)^m - n^m$, and  the probability that an injective relation $R \subseteq Y \times X$ selected uniformly at random has a fixed point is $1 - ( \frac{n}{n+1} )^m$.

Finally, we want to mention some unsolved problems,
which are related to the topic of the present paper and
yield material for future work.

In Section \ref{WITHOUTFP} we have shown that on an $n$-set $X$
the probability of a randomly selected involution on $X$ to 
be proper tends to $0$ if $n$ tends to infinity.
We conjecture that also the probability of a randomly selected 
permutation on $X$ to be an involution tends to $0$ if 
$n$ tends to infinity.
But a proof is still missing.

Concerning the three classes of Section \ref{IDEMP}, we hope to 
be able to show for the (as we believe) simplest case 
$\FP^i_n(0),\ldots, \FP^i_n(n)$ how these numbers are
related to each other with respect to their sizes.
So far, we only know that 
$k \leq \sqrt{\frac{5}{4} + n} - \frac{3}{2}$ implies
$\FP^i_n(k) \leq \FP^i_n(k+1)$.
Unfortunately, this is not an equivalence.
For $n = 5$ we get $\sqrt{\frac{5}{4} + n} - \frac{3}{2} = 1$,
such that
$0 = \FP^i_5(0) \leq \FP^i_5(1) \leq \FP^i_5(2) = 80$.
But it holds $\FP^i_5(3) = 90$.
Only then the numbers are decreasing as
$\FP^i_5(4) = 20$ and $\FP^i_5(5) = 1$.

An interesting property of a relation $R$ on a set $X$
is to have a kernel, which is a subset $K$ of $X$ such that
$x \notin K$ iff there exists $y \in K$ with $(x,y) \in R$,
for all $x \in X$.
It can be shown that in case $|X| = n \geq 1$ the probability that 
a randomly selected relation on $X$ has a kernel is at most $S(n)$, 
where the 
function $S : \NAT_{n \geq 0} \to \REAL$ is defined by
$S(n) = \sum\limits_{i=0}^n \binom{n}{i} \frac{(2^i - 1)^{n-i}}{2^{ni}}$.
Due to the \RelView-experiments described in \cite{BerKernel} we 
conjecture that $\lim\limits_{n \rightarrow \infty} S(n) = 0$.
This means that, given a finite and non-empty set $X$, the probability 
of a randomly selected relation on $X$ to have a kernel tends to $0$ 
if $|X|$ tends to infinity.
Our conjecture is supported by computing $S(n)$ up to $n = 4001$ 
by a C-program together with the GNU MP library for arbitrary precision 
arithmetic on integers and rational numbers. 
The results show that the sequence $(S(n))_{n \geq 1}$ decreases from
$S(1) = 0.5$ down to $S(4001) = 0.139942.$
However, they also show that it alternates between being 
decreasing and being increasing, 
i.e., has a wave-like behavior. 
E.g., from $S(1)$ to the first local minimum $S(311) = 0.193723$ it 
strictly decreases.
Then the first wave starts and $(S(n))_{n \geq 1}$ 
strictly increases up to the first local maximum $S(383) = 0.195389$ 
and then strictly decreases to the second local minimum $S(686) = 0.169429$.
Looking at the subsequences of the local maximums and minimums of 
$(S(n))_{n \geq 1}$ it seems that both are strictly decreasing, which
additionally supports our conjecture. 
In future work we would like to verify this conjecture.

We would like to thank the anonymous reviewers for taking the time and effort necessary to review 
the current paper. We sincerely appreciate all valuable comments and suggestions, 
which helped us to improve and extend the content of this paper.



\section*{Appendix}

\noindent
\textbf{Proof of Lemma \ref{BINOM}.}
\begin{enumerate}[(a)]
\item The claim is shown by the calculation
$$
  \binom{n}{k+1} = \frac{n!}{(k+1)!(n-(k+1))!} 
= \frac{(n-k)n!}{(k+1)k!(n-k)!} 
= \frac{n-k}{k+1} \binom{n}{k}
$$
that applies the definition of the binomial coefficient stated in Section~\ref{PRELIM}.
\item Again the claim follows immediately from
\[
\binom{n+1}{k} = \frac{(n+1)!}{k!(n+1-k)!} 
= \frac{(n+1)n!}{(n+1-k)k!(n-k)!} 
= \frac{n+1}{n+1-k} \binom{n}{k}.
\]
\item This follows from (a) and (b).
\end{enumerate}

\bigskip

\noindent
\textbf{Proof of Lemma \ref{LEMMW-1}.}
We use induction on $n$.
The induction base $n = 8$ follows from $8^7 = 2\,097\,152$ and
the following nine equations:
$$
\begin{array}{r@{~}c@{~}l@{\qquad}r@{~}c@{~}l@{\qquad}r@{~}c@{~}l}
2^8 0^8 2^0 &=& 0 &
2^8 1^7 2^1 &=& 512 &
2^8 2^6 2^2 &=& 65\,536 \\
2^8 3^5 2^3 &=& 497\,664 &
2^8 4^4 2^4 &=& 1\,048\,576 &
2^8 5^3 2^5 &=& 1\,024\,000 \\
2^8 6^2 2^6 &=& 589\,824 &
2^8 7^1 2^7 &=& 229\,376 &
2^8 8^0 2^8 &=& 65\,536 
\end{array}
$$
For the induction step, assume $n \in \NAT$ such that $n \geq 8$
and the induction hypothesis for $n$ to be true.
Furthermore, let an arbitrary $k \in \NAT$ with $k \leq n+1$ be given.
If $k < n+1$, then we have:
\begin{align*}
2^{n+1} k^{n+1-k} 2^k &= 2k 2^n k^{n-k} 2^k  \\
                      &\leq 2k n^{n-1} & \mbox{induction hypothesis} \\
                      &\leq 2 n^n & \mbox{as } k \leq n\\
                      &\leq (n+1)^n & \mbox{as } 2 \leq (1+\frac{1}{n})^n
\end{align*}
The remaining case $k = n+1$ is shown as follows:
\begin{align*}
2^{n+1} k^{n+1-k} 2^k &= 2^{n+1} 2^{n+1} & \mbox{as } k^0 = 1 \\
                      &= 4 \cdot 2^n n^{n-n} 2^n & \mbox{as } n^0 = 1 \\
                      &\leq 4 n^{n-1} & \mbox{induction hypothesis for $k=n$} \\
                      &\leq n^n & \mbox{as }4\leq 8\leq n\\
                      &\leq (n+1)^n
\end{align*}

\noindent
\textbf{Proof of Lemma \ref{LEM2}.}
(a)
First we show that $e^{W(n{\mathrm e})-1}=\frac{n}{W(n{\mathrm e})}$.
Starting with $\frac{n}{W(n{\mathrm e})}$ we get:
\begin{align*}
\frac{n}{W(n{\mathrm e})}
&= \frac{n{\mathrm e}}{{\mathrm e}W(n{\mathrm e})}\\
&= \frac{W(n{\mathrm e}){\mathrm e}^{W(n{\mathrm e})}}
        {{\mathrm e}W(n{\mathrm e})} 
   && \mbox{definition $W$}\\
&= \frac{{\mathrm e}^{W(n{\mathrm e})}}{{\mathrm e}}\\
&= {\mathrm e}^{W(n{\mathrm e})-1}
\end{align*}
Now let $k_n:={\mathrm e}^{W(n{\mathrm e})-1}=\frac{n}{W(n{\mathrm e})}$. For the derivative $\frac{\d h_n}{\d k}$ of 
$h_n(k) = k^{n-k}$ we calculate
$$\frac{\d h_n}{\d k}(k) = (n-k)k^{n-k-1}-k^{n-k}\ln(k)
                         = k^{n-k-1}(n-k-k\ln(k))
$$ 
(see~\cite{Wiki-differentiation}) so that 
\[
\frac{\d h_n}{\d k}(k) = 0 \iff n = k + k \ln(k)=k(1+\ln(k)).
\]
Next, we calculate as follows:
\begin{align*}
n=k(1+\ln(k))
&\iff n{\mathrm e}=k{\mathrm e}(1+\ln(k))\\
&\iff n{\mathrm e}={\mathrm e}^{1+\ln(k)}(1+\ln(k))\\	
&\iff W(n{\mathrm e})=1+\ln(k) 
      && \mbox{definition $W$}\\
&\iff k=e^{W(n{\mathrm e})-1}
\end{align*}

Hence $\frac{\d h_n}{\d k}(k) = 0 \iff k = k_n$.
Since $k(1+\ln(k))$ is monotonic in $k$,
$\frac{\d h_n}{\d k}(k) \geq 0
\iff n \geq k(1+\ln(k))
\iff k \leq k_n$,
and thus $h_n$
is increasing in the interval $[1,k_n]$
(since $k \geq 1$ is assumed).
Similarly,
$\frac{\d h_n}{\d k}(k) \leq 0
\iff n \leq k(1+\ln(k))
\iff k \geq k_n$,
and thus $h_n$
is decreasing for $k \geq k_n$.
Hence $k_n$ is the maximal value of $h_n$.

(b) 
We show that $2\lfloor 3\log_2(n)\rfloor+1\leq\frac{n}{W(n{\mathrm e})}$,
so that the original assertion follows from (a). 
First note that for all $x, y \in \REAL_{\geq 0}$ we have 
$W(x) \leq y$ iff $x \leq y{\rm e}^y$, which follows from 
the definition of $W$ and the fact that $W$ is increasing 
(see \cite{WikiLambert}). 
Using the abbreviation $m=2\lfloor 3\log_2(n)\rfloor+1$ we obtain
\begin{align*}
	m\leq\frac{n}{W(n{\mathrm e})}
	&\iff W(n{\mathrm e})\leq\frac{n}{m}\\
	&\iff n{\mathrm e}\leq \frac{n}{m}{\mathrm e}^{\frac{n}{m}}\\			
	&\iff 1\leq \frac{1}{m}{\mathrm e}^{\frac{n}{m}-1}\\
	&\iff 1\leq {\mathrm e}^{\frac{n}{m}-1-\ln(m)}\\
	&\iff 0\leq \frac{n}{m}-1-\ln(m)\\			
	&\iff (\ln(m)+1)m\leq n.
\end{align*}
Now, we use the definition of $m$ and then the inequality 
above follows from
\[
(\ln(m)+1)m\leq 2^{\frac{m-1}{6}}
\]
since 
$$
2^{\frac{m-1}{6}}
  = (2^{\frac{m-1}{2}})^\frac{1}{3}
  = (2^{\lfloor 3\log_2(n)\rfloor})^\frac{1}{3}
  = (2^{\lfloor\log_2(n^3)\rfloor})^\frac{1}{3}
  \leq (n^3)^\frac{1}{3}
  = n. 
$$
This estimate is true for all $m \geq 49$ ($=$ holds at approximately 
$48.275$). 
Consequently, the original inequality holds if
$2\lfloor 3\log_2(n)\rfloor+1\geq 49$, 
which is equivalent to $\lfloor 3\log_2(n)\rfloor\geq 24$. 
This is implied by $n^3\geq 2^{24}$ and, hence, by $n\geq 2^8=256$.

\bigskip

\noindent
Note that part (b) of Lemma \ref{LEM2} already holds for $n \geq 228$, 
which can be tested by computing the corresponding values for 
$n \in \{228,\ldots,255\}$. 
This stronger result is not needed in the current paper.
\end{document}